\documentclass[fleqn,usenatbib]{mnras}

\usepackage{newtxtext,newtxmath}

\usepackage[T1]{fontenc}

\DeclareRobustCommand{\VAN}[3]{#2}
\let\VANthebibliography\thebibliography
\def\thebibliography{\DeclareRobustCommand{\VAN}[3]{##3}\VANthebibliography}

\usepackage{graphicx}	
\usepackage{amsmath}	

\usepackage{graphicx}	
\usepackage{amsmath}	
\usepackage{tabularx}
\usepackage[table,xcdraw]{xcolor}
\usepackage{newtxtext,newtxmath}
\usepackage{float}

\newcommand\HI{${\rm H}${\sc i}}

\newcommand\Htwo {${\rm H}_{2}$}
\newcommand{\HH}{H$_2$}

\newcommand\eagle{{\sc eagle}}       
\newcommand\gadget{{\sc gadget}}       
\newcommand\subfind{{\sc subfind}}       
\newcommand\orbweaver{{\sc OrbWeaver}} 
\newcommand\mstar{$\log(M_{\star}/{\rm M_{\odot}})\ $}       
\newcommand\mhalo{$\log(M_{\rm 200}/{\rm M_{\odot}})\ $}
\title[The balance of gas flows in \eagle\ satellites]{An orbital perspective on the starvation, stripping, and quenching of
satellite galaxies in the \eagle\ simulations}

\author[R. J. Wright et al.]{Ruby J. Wright\thanks{E-mail: ruby.wright@icrar.org}$^{ 1,2}$, Claudia del P. Lagos$^{1,2}$, Chris Power$^{1,2}$, Adam R. H. Stevens$^{1,2}$, \newauthor{Luca Cortese$^{1,2}$, Rhys J. J. Poulton$^{1,2}$} \\\ \\ 
$^{1}$International Centre for Radio Astronomy Research (ICRAR), University of Western Australia, Crawley, WA 6009, Australia\\
$^{2}$ARC Centre of Excellence for All Sky Astrophysics in 3 Dimensions (ASTRO 3D)}

\date{Accepted XXX. Received YYY; in original form ZZZ}

\pubyear{2022}

\begin{document}
\label{firstpage}
\pagerange{\pageref{firstpage}--\pageref{lastpage}}
\maketitle

\begin{abstract}
Using the \eagle\ suite of simulations, we demonstrate that both cold gas stripping {\it and} starvation of gas inflow play an important role in quenching satellite galaxies across a range of stellar and halo masses, $M_{\star}$ and $M_{200}$. Quantifying the balance between gas inflows, outflows, and star formation rates, we show that even at $z=2$, only $\approx30\%$ of satellite galaxies are able to maintain equilibrium or grow their reservoir of cool gas -- compared to $\approx50\%$ of central galaxies at this redshift. We find that the number of orbits completed by a satellite on first-infall to a group environment is a very good predictor of its quenching, even more so than the time since infall. On average, we show that intermediate-mass satellites with $M_{\star}$ between $10^{9}{\rm M}_{\odot}-10^{10}{\rm M}_{\odot}$ will be quenched at first pericenter in massive group environments, $M_{200}>10^{13.5}{\rm M}_{\odot}$; and will be quenched at second pericenter in less massive group environments, $M_{200}<10^{13.5}{\rm M}_{\odot}$. On average, more massive satellites ($M_{\star}>10^{10}{\rm M}_{\odot}$) experience longer depletion time-scales, being quenched between first and second pericenters in massive groups; while in smaller group environments, just $\approx30\%$ will be quenched even after two orbits. Our results suggest that while starvation alone may be enough to slowly quench satellite galaxies, direct gas stripping, particularly at pericenters, is required to produce the short quenching time-scales exhibited in the simulation.

\end{abstract}


\section{Introduction}\label{sec:introduction}
Understanding how, when, and why star formation is observed to cease in galaxies remains a topic of great interest in extragalactic astronomy. The mechanisms that quench galaxies must reduce the amount of cool gas available for star formation, which can be achieved by either directly removing gas from galaxies, and/or stifling the rate at which the gas is replenished via fresh accretion. In the case of satellite galaxies, the picture is further confounded by the fact that processes both internal and external to galaxies are known to act, and can be responsible for simultaneous gas removal (via ejective feedback or stripping) {\it and} reduced gas inflow (via preventative feedback or starvation). The interplay between these different processes, and how they cause the eventual quenching of satellite galaxies, remains poorly understood in the literature.

The evolution of a galaxy is intrinsically tied to the continuous balance between baryonic processes in the inter-stellar medium (ISM), circum-galactic medium (CGM), and surrounding larger scale environment. Based on this idea, commonly adopted ``gas-regulator'' equilibrium models appeal to continuity arguments to claim that there must be a tightly maintained balance between gas replenishment, star formation, and gas removal in galaxies (e.g. \citealt{Oort1970,Larson1972,Tinsley1980,White1991,Keres2005,Bouche2010,Lilly2013,Dekel2014}).  Such models can be simplistically represented as in Equation \ref{eq:isminst}:

\begin{equation}\label{eq:isminst}
\dot{M}_{\rm gas}=\dot{M}_{\rm in}-\dot{M}_{\rm out}-\dot{M}_{\star},
\end{equation}
\noindent{where $\dot{M}_{\rm gas}$ is the net rate of change in mass of a galaxy's gas reservoir, $\dot{M}_{\rm in}$ is the gas inflow/accretion rate, $\dot{M}_{\rm out}$ is the rate of gas removal, and $\dot{M}_{\star}$ is the conversion rate of gas to stars.\footnote{We remark that this equation neglects the return of gas to the ISM via stellar winds, which have a relatively small effect on total gas masses.}} 

 It is clear that an imbalance in any of the processes constituting the baryon cycle directly leads to the modulation of another. One readily-observed property that clearly reflects the balance of baryonic processes in a galaxy is its star formation rate (SFR) -- either measured directly, or via colour as a proxy. Many large observational studies have proven the existence of a bi-modality in the star-formation status of galaxies at low redshift (e.g. \citealt{Strateva2001,Blanton2003,Kauffmann2003,Kauffmann2004,Baldry2004,Balogh2004,Brinchmann2004,Baldry2006,Wyder2007,Taylor2015,Davies2016,Thorne2021,Katsianis2021}). The existence of a passive population, characterised by ``quenched'' star formation, suggests that the gas capable of forming stars in these galaxies have been depleted. Fundamentally, this depletion must be due to an imbalance in the rate of gas replenishment and the rate of gas removal -- that is to say, gas inflow rates are inadequate to offset the combined depletion or removal of gas via star formation and outflows (driven by internal and/or external mechanisms). 

The physical mechanisms responsible for galaxy-scale gas depletion have long been argued to be strongly dependent on environment. Previous studies have shown that when controlled for stellar and halo mass, satellite galaxies exhibit systematically higher quenched fractions compared to central galaxies -- particularly at stellar masses below \mstar$\approx10$ (e.g. \citealt{Balogh1999,Gomez2003,Cooper2008b,vandenBosch2008,Weinmann2009,Peng2010,Peng2012,Wetzel2012,Wetzel2013,Knobel2013,Robotham2014,Grootes2017,Treyer2018,Davies2019b}). The documented increase in the quenched fraction of satellites has also been linked with a deficiency of cold gas, as statistically demonstrated in both observations (e.g. \citealt{Giovanelli1985,Brown2017}) and hydrodynamical simulations (e.g. \citealt{Marasco2016,Stevens2019}).

\citet{Cortese2021} categorise the different physical mechanisms behind the quenching of satellites into 4  groups: 
 \begin{enumerate}
     \item {\bf starvation} -- lack of gas inflow;
     \item{\bf environmental stripping} --  direct cool gas removal;
     \item {\bf internal ejective feedback} --  direct cool gas removal;  
     \item {\bf reduced star formation efficiency} (SFE). 
 \end{enumerate}
 
Mechanisms (i) -- (iii) involve a reduction in total galaxy gas mass, while mechanism (iv) reduces the efficiency with which gas, even if present, can be converted to stars. A reduction in SFE is predicted in a number of scenarios, for instance morphological quenching (e.g. \citealt{Martig2009}), unfavourable ISM metallicity \& radiation conditions (e.g. \citealt{Krumholz2009}), or where there is adequate support of the gas against collapse (e.g. \citealt{Peng2020}). For the purposes of this paper, we focus on the quenching mechanisms that actively reduce the amount of cool gas in galaxies -- i.e. mechanisms (i), (ii), \& (iii) -- which combine to produce the observed deficiency of gas in satellite galaxies mentioned above. As such, we treat mechanism (iv) as a potentially important, but second-order effect \citep{Davis2014}. 
 
 While each of the categories mentioned above can be considered independent physical mechanisms, it is generally agreed that several must act simultaneously to quench satellite galaxies. {\it Starvation} is a general term that refers to a reduction or cessation of gas accretion onto galaxies (e.g. \citealt{Larson1980}). Even in the most extreme cases of gas stripping, adequate gas inflow to replenish the ISM of satellites could theoretically sustain star formation -- and as such, it is posited that the full quenching of satellites requires some degree of starvation to occur. The role of starvation in the quenching of satellites is  hard to constrain due to the difficulties associated with directly measuring gas inflows. Hydrodynamical simulations, with their ability to self-consistently model the behaviour of gas over vast physical scales, have thus been a very useful tool to probe influence of gas accretion (or, lack thereof) on galaxy evolution. 
 
 A reduction in gas accretion to galaxies can arise at different scales: from either a lack of gas inflow from the IGM to the halo of a galaxy, or due to an inability of CGM gas to cool onto the galactic disk. In hydrodynamical simulations, the preventative impact of stellar and AGN feedback on both halo- and galaxy-scale gas accretion has been clearly demonstrated in central galaxies (e.g. \citealt{Voort2011,FaucherGiguere2011,Nelson2015,Correa2018a,Correa2018b,Mitchell2020b,Wright2020}).
 
 In the case of satellites, there is the additional consideration that gas can be stripped from the hot halo of galaxies on infall due to interaction with the intra-group/intra-cluster medium of their host. Since this hot gas stripping induces a reduction of cold gas inflow at the scale of the ISM, we still refer to this as a starvation process as opposed to a gas removal process (commonly referred to as ``strangulation'', e.g.  \citealt{Balogh2000b}). In satellites, the distinction between {\it starvation} and {\it stripping} is simply a choice of scale. We adopt the convention that any process reducing gas accretion {\it at the scale of the galactic disk} is a starvation process (including the ``strangulation'' scenario).\footnote{In any case, at a larger scale \citet{Wright2020} show that gas accretion to satellite sub-haloes is strongly suppressed in group environments (Fig. 7).} Furthermore, stellar and AGN feedback have been shown to actually increase the efficiency of these hydrodynamical gas removal mechanisms in  simulations (e.g. \citealt{Bahe2015,Zoldan2017}). Many semi-analytic models (SAMs) assume this strangulation scenario for satellites, in which all hot gas surrounding satellites is completely and immediately stripped upon infall (e.g. \citealt{Cole2000,Somerville2008,Lagos2018a}). 

It is also important to consider the direct {\it stripping} of cold gas in satellite galaxies, and how this influences their evolution. We note that we use the term ``cold gas'' to collectively refer to a galaxy's combined reservoir of neutral gas (both atomic, \HI; and molecular, \Htwo). Additionally, while this stripping can either be hydrodynamical (e.g. ram pressure stripping; \citealt{Gunn1972}) or gravitational in nature (e.g. tidal stripping; \citealt{Boselli2006}); we do not focus on this distinction -- and use the term {\it cold gas stripping} to encompass all environmentally-driven cold gas removal processes. Observational evidence of satellite cold gas stripping is plentiful in the cluster environment (e.g. \citealt{Fossati2018,Jachym2019,Moretti2020}). Stripping of cold gas from satellites is also ubiquitously predicted in hydrodynamical simulations, both in cosmological (e.g. \citealt{Bahe2015,Lotz2019}) and idealised (e.g. \citealt{Tonnesen2009,Bekki2014, Steinhauser2016}) cases. Semi-analytic studies have also highlighted that including a physical model for cold-gas stripping, in addition to the canonical satellite strangulation model, may better reproduce observed quenched fractions and \Htwo-\HI\ ratios (e.g. \citealt{Stevens2017,Cora2018,Xie2020}). While it is clear that cold gas stripping is important, the magnitude of its role in quenching satellites -- and how this varies with stellar and halo mass -- is still debated \citep{Cortese2021,Oman2021}. 

Focusing on the {\it starvation} of satellites, \citet{Voort2017} studied the modulation of cool gas inflow rates onto galaxies as a function of stellar and halo mass in the \eagle\ simulations. They find strong environmentally-driven suppression of gas accretion onto satellites in groups above a mass of \mhalo$\approx12.5$, and show that this suppression -- even in the absence of direct gas stripping -- could be sufficient to completely explain the quenching of low-mass satellites. This is also the preferred method of quenching satellites in several SAMs (e.g. \citealt{Lagos2018a,Baugh2019}). The question becomes one of quenching time-scale -- a galaxy experiencing starvation would be expected to slowly use up its available gas reservoir on a gentle path towards quiescence; compared to a galaxy that is being simultaneously starved and directly stripped of cold gas that would be expected to quench rapidly.
 
Quenching time-scales -- the time interval taken for a galaxy to transform from star forming to quiescent -- can be measured using a wide variety of methodologies, particularly in the case of satellites (see \citealt{Cortese2021} for a comprehensive review). A popular method for measuring satellite quenching time-scales in observations uses a  ``delayed-then-rapid'' parameterisation, whereby satellites continue forming stars for some time interval after infall to a group, before experiencing a rapid reduction in star formation activity \citep{DeLucia2012}. This approach has lead to consistent measurements of quenching time-scales in excess of $\approx2$ Gyr, though exact time-scales depend on the particular study (e.g. \citealt{Wetzel2013,Hirschmann2014,Taranu2014,Haines2015,Oman2016,Oman2021}). These ``long'' quenching time-scales have typically been considered most compatible with a starvation-like scenario; while it is normally argued that quenching time-scales of $<1$ Gyr reflect a stripping-dominated scenario (as determined from studies of cluster satellites, e.g.  \citealt{Boselli2016,Fossati2018}). In the \eagle\ simulations, \citet{Wright2019} show that the quenching time-scales of satellites exhibit a large spread, with satellites posessing low stellar--halo mass ratios ($M_{\star}/M_{200}<10^{-4}$) typically quenching very quickly ($\ll1$ Gyr), compared to those with higher stellar--halo mass ratios ($M_{\star}/M_{200}>10^{-3}$) that show more protracted quenching time-scales ($>2$ Gyr). 

A question related to quenching time-scales is whether satellites are expected to quench by their first pericentric passage. Different approaches have been found to yield different answers to this question, though it is generally agreed that there is a dependence on satellite and host halo mass. Idealised simulations have shown that satellites with a stellar mass of \mstar$>10$ are likely to be significantly stripped at first pericenter in a cluster environment, but not completely quenched (e.g. \citealt{Steinhauser2016}). Interestingly, the picture for quenching appears more dramatic in larger-scale cosmological simulations. \citet{Lotz2019} show that over $90\%$ of satellites are quenched after first pericenter in the {\sc magneticum} simulations, with only extremely massive satellites, \mstar$>11$, exhibiting continuing star formation (see also \citealt{Bahe2015}). We note that limited particle resolution in these large-scale simulations -- and the associated lack of a true ``cold'' phase -- may lead to an over-prediction in the forces associated with stripping (e.g. \citealt{Bahe2017b}). 

While it is clear that satellites experience environmentally-driven quenching, and that both starvation and cold gas stripping play a role, the relative influence of each -- as a function of satellite mass, host mass, and orbital time -- remains poorly understood. In this study, we use the \eagle\ simulations to provide a comprehensive analysis of the balance between gas inflows, outflows, and star formation in satellite galaxies, and how this balance varies {\it along their orbits}. In doing so, we aim to answer the question: under what circumstances can satellites continue to accrete gas, and continue to form stars?

This paper is arranged as follows: in \S\ref{sec:methods}, we outline the \eagle\ simulation suite and structure finding algorithm (\S\ref{sec:methods:eagle} \& \ref{sec:methods:subfind}); our methodology to define the ISM of galaxies and relevant gas flow rates (\S\ref{sec:methods:gasflow}); and the \orbweaver\ tool used to analyse the orbits of satellites in the simulation (\S\ref{sec:methods:orbweaver}). In  \S\ref{sec:results:mhalo}, we analyse how the balance of gas flows in satellites varies as a function of stellar mass, halo mass, and redshift. In  \S\ref{sec:results:orbits}, we analyse how the balance of gas flows in satellites varies explicitly as a function of orbital time. Finally, in \S\ref{sec:conclusions} we conclude what our findings broadly tell us about the evolution of satellite galaxies, and outline future research direction.

\newpage
\section{Methods}\label{sec:methods}
\subsection{The \eagle\ simulations}\label{sec:methods:eagle}

The \eagle\ (Evolution and Assembly of GaLaxies and their Environments) simulation suite \citep{Schaye2015,Crain2015} is a collection of cosmological hydrodynamical simulations that self-consistently model the co-evolution of baryonic and dark matter (DM) in a cosmological context. \eagle\ uses the  the \gadget-3 tree-SPH (smoothed-particle hydrodynamics) code scheme outlined in \citet{Springel2005}, together with the {\sc anarchy} set of revisions outlined in \citet{Schaller2015} to track the formation and evolution of galaxies down to $z=0$ in a collection of periodic volumes with varying sub-grid physics. In the main body of this paper, we make use of the $100$ Mpc intermediate-resolution \eagle\ run, with key characteristics listed in Table \ref{tab:eagle}. 

\eagle\ adopts the parameters\footnote{As per Table 1 in \citet{Schaye2015}: $\Omega_{\rm m}=0.307$, $\Omega_{\Lambda}=0.693$, $\Omega_{\rm b}=0.04825$, $h\equiv H_{0}/(100\ {\rm km s^{-1} Mpc^{-1}})=0.6777$, $\sigma_{8}=0.8288$, $n_{\rm s}=0.9611$, $Y=0.248$.} of a ${\Lambda}$CDM universe from \citet{PlanckCollaboration2014}, with initial conditions generated using the method outlined in \citet{Jenkins2013}. Sub-grid models are required to track otherwise unresolved (sub-kpc) physical processes relevant to galaxy formation and evolution. This includes consideration of radiative cooling and photoheating, star formation, stellar evolution and metal enrichment, stellar feedback, and super-massive black hole (SMBH) growth and feedback (active galactic nuclei; AGN). These sub-grid models introduce a number of free parameters, each calibrated to match $z\approx0$ observations of either the galaxy stellar mass function, the galaxy size--mass relation, or the galaxy SMBH mass--stellar mass relation.

Photoheating and radiative cooling are implemented in \eagle\ based on the work of \citet{Wiersma2009}, including the influence of 11 elements: H, He, C, N, O, Ne, Mg, Si, S, Ca, and Fe \citep{Schaller2015}; with the UV and X-ray background described by \citet{Haardt2001}. Since the \eagle\ simulations do not provide the resolution to model cold, interstellar gas, a density-dependent temperature floor is imposed (normalised to $T=8\ 000$~K at $n_{\rm H}=10^{-1}{\rm cm}^{-3}$). To model star formation, a metallicity-dependent density threshold is set, above which star formation is locally permitted \citep{Schaye2015}, and gas particles meeting this threshold are converted to star particles stochastically. The star formation rate is set by a tuned pressure law \citep{Schaye2007}, calibrated to the work of \citet{Kennicutt1983} at $z = 0$. 

The stellar feedback sub-grid model in \eagle\ accounts for energy deposition into the ISM from radiation, stellar winds, and supernova explosions. This is implemented based on the prescription outlined in \citet{DallaVecchia2012}, via a stochastic thermal energy injection to gas particles in the form of a fixed temperature boost, $\Delta T_{\rm SF} = 10^{7.5}$K. The average energy injection rate from young stars is given by $f_{\rm th}\times8.73\times10^{15}$ erg per gram of stellar mass formed, assuming a \citet{Chabrier2003} simple stellar population and that $10^{51}$ erg is liberated per supernova event. $f_{\rm th}$ is set by local gas density and metallicity, ranging between $0.3-3$ in the fiducial \eagle\ model. 

SMBHs are seeded in \eagle\ when a halo exceeds a virial mass of $10^{10}\ h^{-1} {\rm M}_{\odot}$, with the seed SMBHs having an initial mass of $10^{5}\ h^{-1} {\rm M}_{\odot}$. Subsequently, SMBHs can grow via Eddington-limited-accretion \citep{Schaye2015}, as well as mergers with other SMBHs, according to work outlined in \citet{Springel2005a}. Similar to stellar feedback, AGN feedback in \eagle\ also involves the injection of thermal energy into surrounding particles in the form of a temperature boost of ${\Delta}T_{\rm AGN}=10^{8.5}$K (in the reference physics run; \citealt{Schaye2015}). The rate of energy injection from AGN feedback is determined using the SMBH accretion rate, and a fixed energy conversion efficiency.

We note that the \eagle\ model has been shown to accurately reproduce observations of the  SFRs and quenched fraction of galaxies as a function of stellar mass and cosmic time (see \citealt{Furlong2015}). \citet{Trayford2016} and \citet{Correa2019} show that satellite galaxies are responsible for the increased fraction of quenched galaxies in \eagle\ below $M_{\star}\approx10^{10}{\rm M}_{\odot}$, however exact comparison of this result to  observations is difficult due to differences in group-finding methodologies (see e.g. \citealt{Davies2019a}). The \eagle\ model has also been shown to accurately reproduce the observed abundance of \HI\ gas in galaxies as a function of stellar mass \citep{Bahe2016,Crain2017,Dave2020}, and critically, how this varies with environment \citep{Marasco2016}. Based on these findings, we conclude that the EAGLE model is well-suited for this study, where we aim to disentangle the mechanisms behind the quenching of satellite galaxies in a representative cosmological volume.

\begin{table}
\centering
\begin{tabular}{lllll}
$L_{\rm box}\ [{\rm Mpc}]$ & $m_{\rm DM}\ [{\rm M}_{\odot}]$ & $m_{\rm gas}\ [{\rm M}_{\odot}]$ & $\epsilon\ [{\rm kpc}]$ \\
\hline \hline
$100$  & $9.7\times 10^{6}$ & $1.8\times 10^{6}$& 0.70\\
\end{tabular}
\caption{Key characteristics and parameters defining the flagship $(100\ {\rm Mpc})^{3}$ \eagle\ run. $L_{\rm box}$ refers to the (comoving) side-length of the periodic simulation volume,  $m_{\rm DM}$ and $m_{\rm gas}$ refer to the starting DM and baryonic particle mass, and $\epsilon$ refers to the Plummer equivalent gravitational smoothing length.} 
\label{tab:eagle}
\end{table}

\subsection{Structure finding with \subfind}\label{sec:methods:subfind}
For the purposes of this work, we employ the outputs of the \subfind\ structure-finding algorithm \citep{Springel2001,Dolag2009}. DM haloes are identified first using a Friends-of-Friends (FoF) algorithm, where DM particles are linked if their separation is less than 20\% of the average inter-particle distance. Baryonic particles are associated with the group that their closest DM particle resides in. \subfind\ then separates individual subhaloes by identifying gravitationally bound sub-structure. Notably for our purposes, \subfind\ identifies (i) the centre of potential for a given subhalo, (ii) the particles associated with a subhalo, and (iii) the status of the subhalo as a central or satellite -- assigning the label of ``central'' to the subhalo that hosts the deepest gravitational potential within the FoF, and ``satellite'' to any remaining substructure. As such, when two subhaloes in the same FoF group are of a similar mass, we note that there is a level of ambiguity in this classification -- which can lead to spurious swapping of central/satellite classification between timesteps (eg. \citealt{Poole2017,Canas2019}). 

\subsection{Defining the ISM and gas flow rates}\label{sec:methods:gasflow}

In order to fairly calculate and compare gas flow rates in \eagle\ galaxies, one must first define what constitutes the ISM of a galaxy in a manner as agnostic as possible to the galaxy's central or satellite classification. Previous approaches  employed to delineate between a galaxy and its surroundings (discussed and compared in Appendix \ref{sec:app:gasflow:prev}) in simulations are diverse, simply reflecting the scientific goals of each respective project. As we aim to consider gas flow rates in satellite galaxies, our ISM definition must not solely rely on considering all gas within a fixed multiple of the halo $R_{\rm 200}$ as some of these methods do -- which applies well for central galaxies, but is not physically motivated in the case of satellites. Additionally, as we desire to compare instantaneous gas inflow and outflow rates along the orbit of \eagle\ satellites, it is a natural choice to use the finely-spaced \eagle\ ``snipshot'' outputs to calculate averaged gas flow rates, as opposed to the coarsely-spaced ``snapshot'' outputs. We outline how we achieve these objectives with our methodology below. 

Instead of using a simple spherical aperture to calculate gas flux, we elect to focus on gas flows to and from the ``cool'' gas reservoir of galaxies by only considering gas below a temperature of $5\times10^{4}{\rm K}$, as well as any star forming gas, as part of the ISM. This choice reflects the gas that can be physically linked to a neutral molecular \citep{Lagos2015} or atomic \citep{Bahe2016} phase, and thus more closely tied to the physics of star formation. For each galaxy, we also impose a spherical boundary defining the upper radial limit for which this cool gas can be included in the ISM. The location of this spherical boundary is determined using the ``BaryMP'' method to fit the baryonic mass profile (BMP) of galaxies \citep{Stevens2014}, outlined fully in Appendix \ref{sec:app:gasflow:ism}. The ISM is defined as any cool gas ($T<5\times10^{4}$K or SFR$>0$) within this radius, and the stellar component as any stellar particles internal to this radius. This ISM selection is illustrated in a group environment in Fig. \ref{fig:s2:bmpgroup}, and an example of the BMP fitting procedure is included in Appendix \ref{sec:app:gasflow:ism}, Fig. \ref{fig:s2:bmpfit}. 

With an ISM definition in hand, one can use a Lagrangian technique to calculate gas flow rates to and from this reservoir.  To calculate gas flow rates, we sum the mass of gas particles that join or leave the ISM between outputs $z_{\rm i}$ and $z_{\rm j}$, and normalise by the time interval between these outputs. Instead of comparing ISM reservoirs between adjacent \eagle\ ``snipshot'' outputs, we compare outputs for gas flow calculations with a  $\Delta t_{\rm ij}$ corresponding to $2$ snipshots; which, at $z=0$, results in a $\Delta t_{\rm ij}$ of $\approx200$ Myr instead of $\approx100$ Myr. This choice allows the gas flow rates to be adequately sampled temporally, while also reducing the amount of galaxies experiencing no gas flux over the time interval. We note, however, that our results are not qualitatively influenced by this selection for any choice of $\Delta t_{\rm ij}$ corresponding to an interval between $1-8$ snipshots. 

We are also able to measure a time averaged SFR in \eagle\ galaxies by summing the mass of all stellar particles within $R_{\rm BMP}$ that were formed between outputs, and normalising by the relevant time interval $\Delta t_{\rm ij}$. This technique allows us to fairly compare inflow, outflow, and star-formation rates in \eagle\ galaxies over a given time interval, which is not possible using the instantaneous particle SFR values. This time-averaged technique also represents a measurement more akin to that in observations \citep{Donnari2019,Donnari2021}.

\begin{figure}
\includegraphics[width=1\columnwidth]{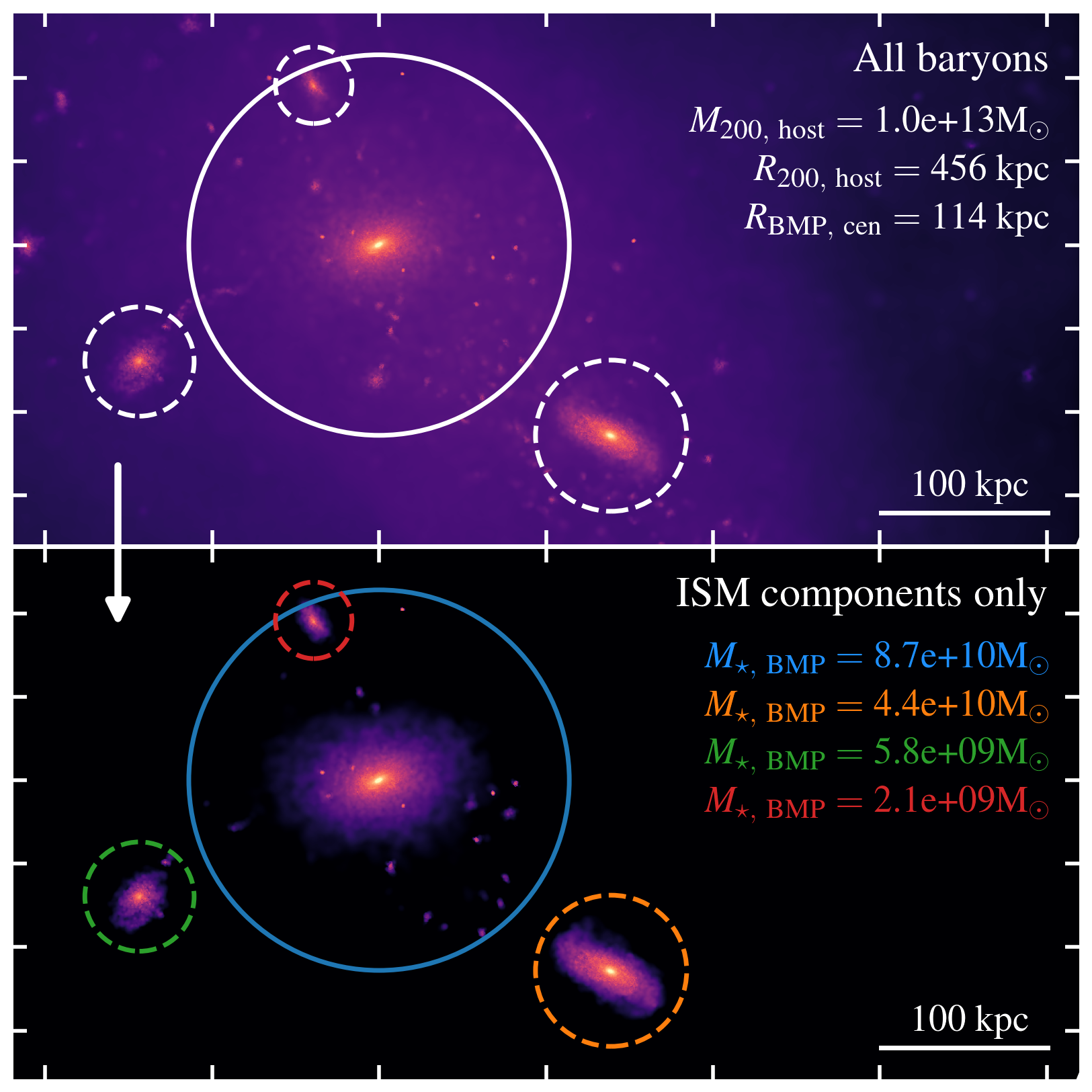}
\caption{An illustration of our ISM selection process for a number of galaxies in a \mhalo$\approx13$ \eagle\ halo at $z=0$. The top panel shows a rendering of all baryons, while the bottom panel solely exhibits the ISM selection. The BMP best-fit radii \citep{Stevens2014} are illustrated with solid lines for the central galaxy, and dashed lines for the surrounding satellites. The stellar masses internal to each  $R_{\rm BMP}$ are quoted in the bottom panel.} 
\label{fig:s2:bmpgroup}
\end{figure}

To summarise, for the remainder of the paper, we use the following terms when referring to different mass reservoirs and gas flow rates:

\begin{itemize}
    \item {\bf Stellar mass ($M_{\star}$)}: the mass of stellar particles within $R_{\rm BMP}$;
    \item {\bf ISM mass ($M_{\rm ISM}$)}: the mass of cool ($T<5\times10^{4}{\rm K}$ or ${\rm SFR}>0$) gas within $R_{\rm BMP}$, approximating a selection of the neutral phase;
    \item {\bf CGM mass  ($M_{\rm CGM}$)}: the total mass of non-ISM gas associated with a \subfind\ (sub)halo; 
    \item {\bf Inflow \& outflow rate} ($\dot{M}_{\rm in}$ \& 
    $\dot{M}_{\rm out}$): the rate of gas mass entering or leaving the ISM, averaged over the interval of 2 snipshots;
    \item {\bf Star formation rate} ($\dot{M}_{\star}$ or SFR): the mass of stars formed within $R_{\rm BMP}$, averaged over the interval of 2 snipshots;
    \item {\bf Halo mass ($M_{\rm 200}$)}: the total mass of a halo within the radius ($R_{\rm 200}$) that encloses an overdensity 200 times greater than the critical density of the universe in the simulation.
    
\end{itemize}

\subsection{Generating an orbital catalogue with \orbweaver}\label{sec:methods:orbweaver}
\textsc{OrbWeaver} is a tool used to extract orbital information from merger trees \citep{Poulton2019,Poulton2020}. It does this by identifying all satellites that come within a factor $N$ of their host virial radius ($R_{\rm vir,\ host}$). Using a root-finding algorithm, the exact timing of the desired $N \times R_{\rm vir,\ host}$ crossings and apsis points are determined. Once these points are found, interpolation is used to calculate the relative satellite-host properties and the associated satellite orbital parameters. Like our gas flow calculations, we ran \textsc{OrbWeaver} on the $200$-output \eagle\ snipshot merger trees to identify and quantify the orbits of all satellites in the simulation. We remark that in this study, we do not focus on galaxies that have been ``pre-processed'' (i.e. those that enter a group/cluster environment already bound to a host). Rather, we seek to understand the physical processes that quench satellites on their first infall to a group environment.


\section{The balance of gas flows in \eagle\ central and satellite galaxies}\label{sec:results:mhalo}

\begin{figure*}
\includegraphics[width=1\textwidth]{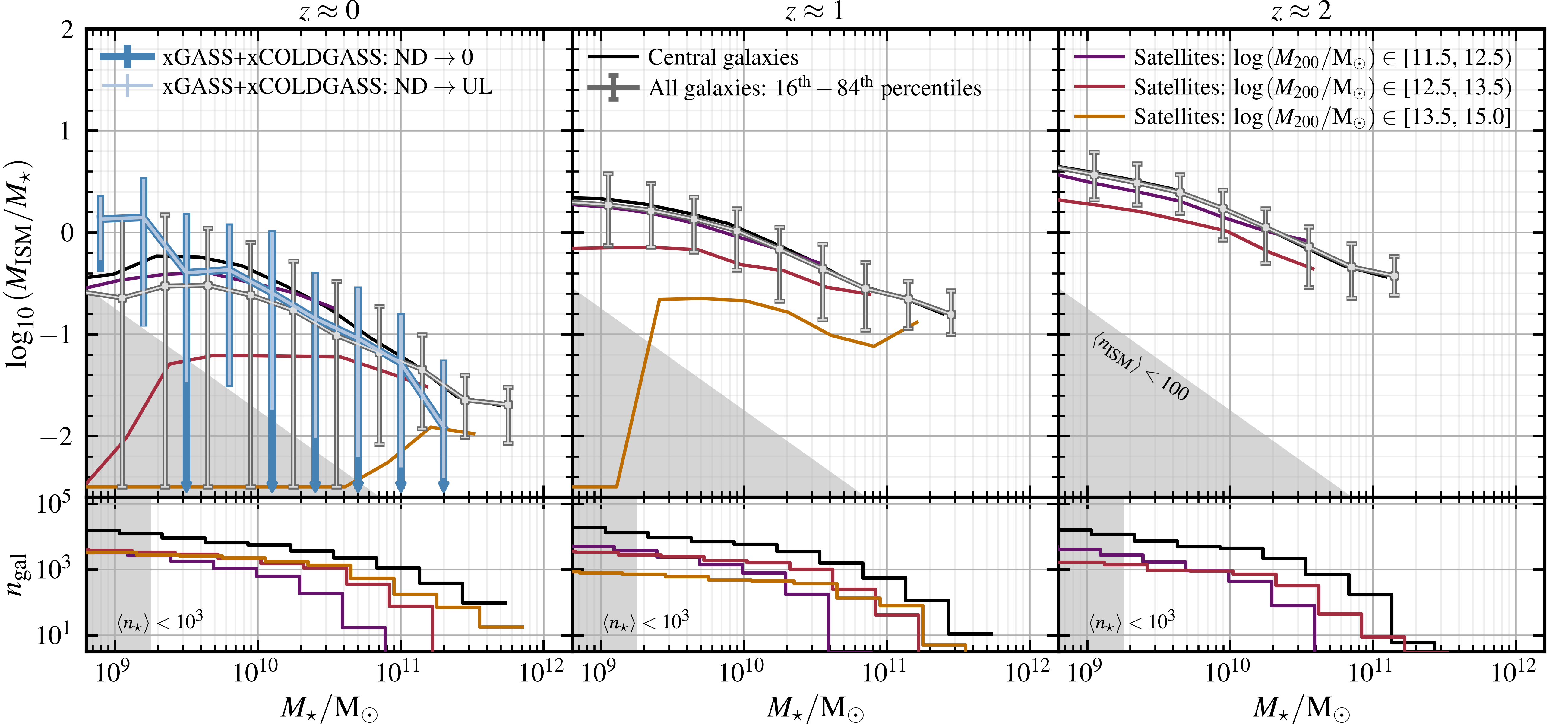}
\caption[The ISM gas mass of \eagle\ centrals and satellites compared to xGASS.]{Comparison of the ISM gas mass ($T<5\times10^{4}{\rm K}$ or ${\rm SFR}>0$) of \eagle\ galaxies, binned by satellite halo mass. These gas masses are shown as a function of stellar mass, with 3 bins per decade in stellar mass. From left to right, results are shown including the $6$ snipshots closest to $z=0$, $z=1$ and $z=2$ respectively. In the left ($z\approx0$) panel, we include a comparison to the xGASS+xCOLDGASS surveys (\citealt{Saintonge2017,Catinella2018}) in blue, with the range indicating $16^{\rm th}$-$84^{\rm th}$ percentiles, with non-detections assumed to be the lower percentile. We convert the summed \HH\ and \HI\ content to equivalent ISM mass by assuming that $M_{\rm H_2}+M_{\rm H I}=M_{\rm ISM}/1.3$, where the factor of $1.3$ accounts for the abundance of Helium. In each panel, galaxies are classified as centrals (black) or satellites split into 3 bins of halo mass: (i) ~sub-$L^{\star}$ haloes with \mhalo$\in [11.5,12.5)$ (purple), (ii) group haloes with \mhalo$\in [12.5,13.5)$ (red), and (iii) larger group/cluster haloes with \mhalo$\in [13.5,15]$ (orange). The medians for all \eagle\ galaxies are also indicated in grey. Error-bars display the $16^{\rm th}-84^{\rm th}$ percentile range in each stellar mass bin. Grey shaded regions in the top panels indicate the ISM mass that would correspond to 100 gas particles. The bottom panels illustrate the number of galaxies included in the top panels for each sample, with the grey regions indicating where the stellar mass of galaxies corresponds to less than $1000$ stellar particles. }
\label{fig:s3:mstar-mgas-mhalo_xgass}
\end{figure*}

In this section, we compare the gas mass, gas flow and star formation rates in central and satellite galaxies as a function of stellar mass and halo mass. In the context of the equilibrium model, the balance between these different different gas flows allows us to predict whether a galaxy is net gaining, maintaining, or losing gas mass from the ISM. While previous work has separately explored inflow and outflow rates in \eagle\ galaxies (e.g. \citealt{Voort2017,Correa2018b,Mitchell2020a,Mitchell2020b})\footnote{While these studies had different science foci to this work, we remark that our measurements of gas inflow and outflow rates in \eagle\ galaxies quantitatively agree with these previous studies to within $0.2$ dex in galaxies with $M_{\star}$ between $10^{9}{\rm M}_{\odot}$ and $10^{12}{\rm M}_{\odot}$.}, we aim to directly compare these flow rates together with star formation rates to explore the evolution of galaxy gas reservoirs as a function of galaxy environment. Specifically, we seek to understand the circumstances under which \eagle\ galaxies {\it net} grow their ISM gas reservoir. 

 \citet{Schaye2015} show that convergence criterion relating to stellar mass functions and star formation rates are not met for \eagle\ galaxies below $\approx 10^{9}{\rm M}_{\odot}$ in stellar mass. As such, we have chosen to remove these galaxies from our analyses in \S\ref{sec:results:mhalo} and \S\ref{sec:results:orbits}.

\begin{figure*}
\includegraphics[width=1\textwidth]{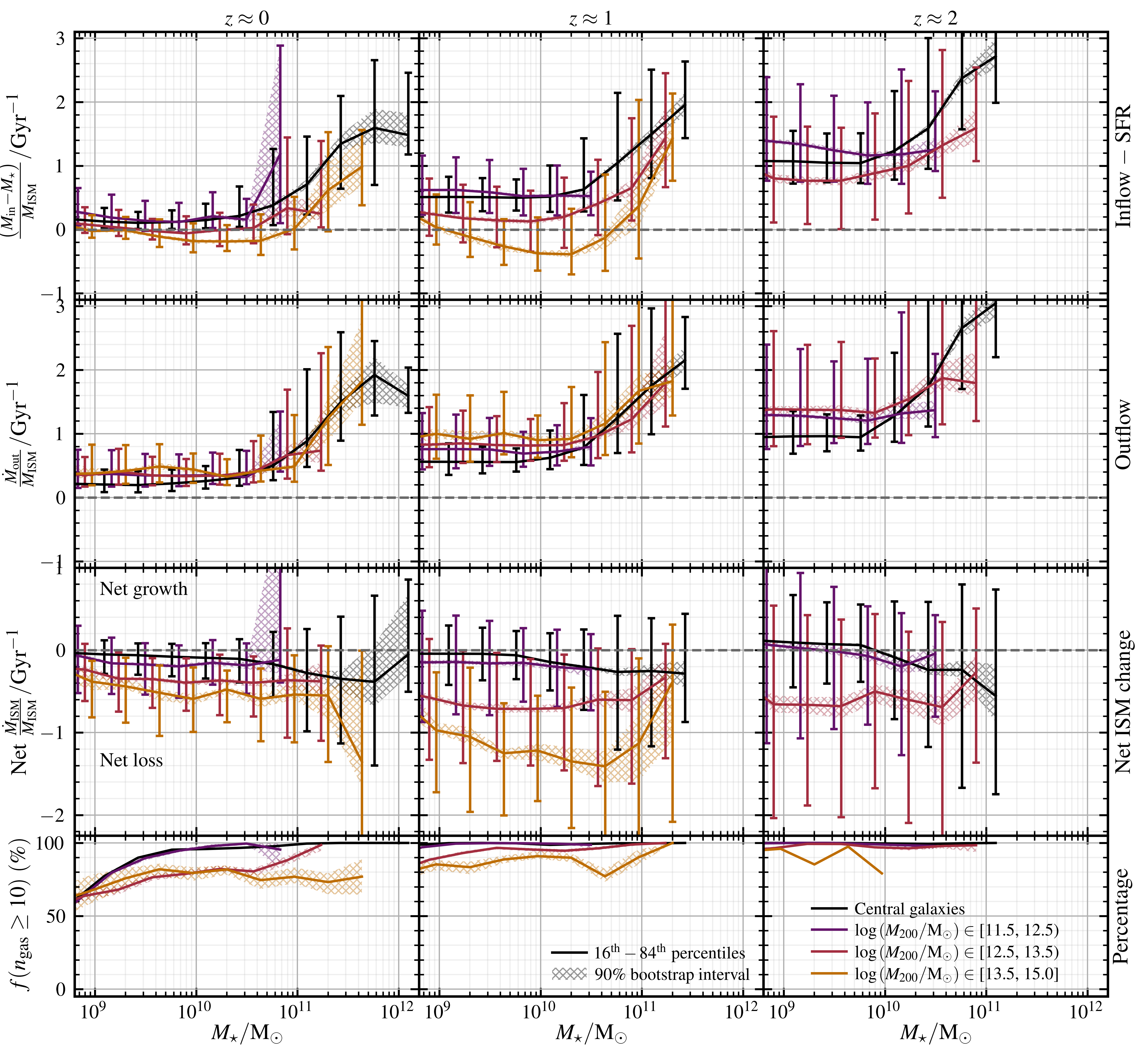}
\caption{ISM inflow subtract star formation rates ($1^{\rm st}$ row), outflow rates ($2^{\rm nd}$ row), and net flow rates ($3^{\rm rd}$ row; sum of $1^{\rm st}$ and $2^{\rm nd}$ rows) in \eagle\ galaxies, normalised by ISM mass. To ensure that the denominator is non-zero and that these specific flow rates are converged, only galaxies with 10 or more ISM gas particles ($M_{\rm ISM}\gtrsim10^{7}{\rm M}_{\odot}$) are included in these calculations. The fraction of galaxies for which this criterion is met is illustrated in the bottom panels. Specific flow rates are shown as a function of stellar mass, with 3 bins per decade in stellar mass. From left to right, results are shown including the $6$ snipshots closest to $z=0$, $z=1$ and $z=2$ respectively. In each panel, galaxies are classified as centrals (black) or satellites split into 3 bins of halo mass: (i) ~sub-$L^{\star}$ haloes with \mhalo$\in [11.5,12.5)$ (purple), (ii) group haloes with \mhalo$\in [12.5,13.5)$ (red), and (iii) larger group/cluster haloes with \mhalo$\in [13.5,15]$ (orange). Error-bars display the $16^{\rm th}-84^{\rm th}$ percentile range in each stellar mass bin, and cross-hatched regions denote the bootstrap-generated $90\%$ confidence interval on the medians (based on 500 re-samples). We only include bins with at least 20 galaxies, and exclude any galaxies undergoing merger events from each of these samples. }
\label{fig:s3:mstar-flowspec-mhalo_z0}
\end{figure*}

Fig. \ref{fig:s3:mstar-mgas-mhalo_xgass} illustrates the ISM gas masses calculated for \eagle\ galaxies at $z\approx0$, $z\approx1$, and $z\approx2$, as per the method outlined in \S\ref{sec:methods:gasflow} and Appendix \ref{sec:app:gasflow}. Each panel contains the results for $5$ samples: one for central galaxies of all halo mass (black), one for all central and satellite galaxies (grey), and 3 samples of satellites split into bins of halo mass: (i) sub-$L^{\star}$ haloes with \mhalo$\in[11.5,12.5)$, (ii) group haloes with \mhalo$\in [12.5,13.5)$  and (iii) larger group/cluster haloes with \mhalo$\in [13.5,15]$. We use a larger bin width for the highest halo mass sample to ensure that we include the most massive haloes in the simulation. 

The $1^{\rm st}$ row of panels in  Fig. \ref{fig:s3:mstar-mgas-mhalo_xgass} illustrates the mass of gas in the ISM of \eagle\ galaxies, as per the definition in \S\ref{sec:methods:gasflow} and Appendix \ref{sec:app:gasflow}. In the left ($z\approx0$) panel, we include a comparison to neutral hydrogen gas content measured the xGASS (extended GALEX Aricebo SDSS Survey; \citealt{Catinella2018}) \& xCOLDGASS (extended CO Legacy Database for GASS; \citealt{Saintonge2017}) surveys, where non-detections have assumed to take the value of the lower percentile. We convert the summed \HH\ and \HI\ content to equivalent ISM mass by assuming that $M_{\rm H_2}+M_{\rm H I}=M_{\rm ISM}/1.3$, where the factor of $1.3$ accounts for the abundance of Helium. We remark that these samples are mass-selected from SDSS, and reside in the redshift range $0.01<z<0.05$. 

Firstly, we note that our measurements of ISM mass (for all galaxies) agree favourably with measurements from the xGASS/xCOLDGASS surveys for galaxies with stellar mass above  $10^{9.5}{\rm M}_{\odot}$. Between $M_{\star}\approx10^{9.5}{\rm M}_{\odot}-10^{11}{\rm M}_{\odot}$, our measurements of gas content are within $0.1-0.2$ dex of the xGASS/xCOLDGASS medians -- giving us confidence in our ISM classification. The two xGASS/xCOLDGASS data-points at stellar mass below $10^{9.5}{\rm M}_{\odot}$ suggest that our measurements of ISM mass may under-predict the gas content of low-mass galaxies relative to observations. We remark that this may be related to numerical issues in \eagle, with galaxies in this mass range having just $\approx100$ gas particles on average.

As demonstrated previously in both observations (e.g. stellar mass -- \HI\ mass relation in \citealt{Brown2017,Calette2021}) and cosmological simulations (e.g. \citealt{Bahe2016,Stevens2019}), we find that at fixed stellar mass, satellites residing in more massive host haloes have lower cool gas content. For satellites in the lowest halo mass sample -- \mhalo$\in [11.5,12.5)$ -- the reduction in gas content relative to centrals is small; only $0.1-0.2$ dex across all stellar mass and redshift. The reduction in gas content in the satellite samples is more dramatic for the small group-mass sample and large group/cluster-mass sample -- with only galaxies above $10^{10.5}{\rm M}_{\odot}$ ($10^{9}{\rm M}_{\odot}$) in the large group/cluster (small group) sample showing ISM gas masses above the floor value at $z\approx0$.

In Fig. \ref{fig:s3:mstar-flowspec-mhalo_z0}, we illustrate the gas flow rates relating to the ISM \eagle\ galaxies (as per \S\ref{sec:methods:gasflow} \& \ref{sec:methods:gasflow}) {\it normalised by ISM mass} as a function of stellar mass. These ``specific'' flow rates, as opposed to total flow rates, reflect the magnitude of influence that the gas flows and star formation will have on the galaxy. For measurements of total inflow rates to \eagle\ satellites, we refer the reader to \citet{Voort2017} -- who find clear environmental suppression of total gas accretion rates to \eagle\ satellites.

The remaining samples of low-mass $z<1$ satellites in groups/clusters reflects those that have not yet been completely stripped of their cool gas, which as we demonstrate in \S\ref{sec:results:orbits}, are likely the satellites that are on first-infall to the host. We stress that the  galaxies remaining in the sample are not atypical, as the vast majority of central galaxies -- reflecting the expected behaviour of satellites pre-infall -- meet this criterion.  We note that the majority of satellite infall events occur between $z\approx0.5$ and $z\approx1.5$ (see \S\ref{sec:results:orbits}), and thus focus a large proportion of our discussion on the $z\approx1$ sample. 

The $1^{\rm st}$ row of panels in Fig. \ref{fig:s3:mstar-flowspec-mhalo_z0} illustrates median ISM inflow rates, subtract SFRs, normalised by ISM mass (which we refer to as ``specific replenishment rate''). This quantity is a measure of the imbalance between star formation and fresh gas accretion — i.e., whether the current rate of gas inflow is adequate to sustain the current star formation rate. At $z\approx1$, the environmentally-driven suppression in gas accretion relative to SFR in satellites becomes evident. This is illustrated in the relative normalisation of the curves for satellites in different halo mass bins, and is particularly notable for galaxies in larger clusters/groups. 

At $z\approx1$, central galaxies maintain a steady median value of $(\dot{M}_{\rm in}-\dot{M}_{\star})/{M_{\rm ISM}}\approx0.3\ {\rm Gyr}^{-1}$ up to \mstar$\approx11$; with satellites in the lowest halo mass sample demonstrating very similar behaviour. Satellites with \mhalo$\in[12.5, 13.5)$ show specific replenishment rates offset negatively from centrals by $\approx0.2$ ${\rm Gyr}^{-1}$, maintaining a value of $(\dot{M}_{\rm in}-\dot{M}_{\star})/{M_{\rm ISM}}\approx 0.1\ {\rm Gyr}^{-1}$ up to \mstar$\approx10.5$. While it is clear that suppression of gas accretion rates occurs in \mhalo$\in[12.5,13.5)$ satellites relative to central galaxies, given that this median value remains positive, we conclude that the average satellite in this sample could not quench in the absence of outflows.

Such is not the case for satellites in the large group/cluster sample, \mhalo$\in[13.5,15]$. These satellites show a clear dip in replenishment rates between \mstar$=9-11$; with a value of $(\dot{M}_{\rm in}-\dot{M}_{\star})/{M_{\rm ISM}}\approx -0.5\ {\rm Gyr}^{-1}$ at \mstar$\approx10$. Assuming that these specific flow rates remain constant, full ISM depletion of the average satellite in this sample would occur in $\approx2$ Gyr -- even in the absence of outflows. We note that the magnitude of suppression of specific replenishment rates in $z\approx0$ satellites relative to $z\approx1$ galaxies is roughly halved, at $(\dot{M}_{\rm in}-\dot{M}_{\star})/{M_{\rm ISM}}\approx -0.25\ {\rm Gyr}^{-1}$ in the most extreme cases.

The $2^{\rm nd}$ row of panels in Fig. \ref{fig:s3:mstar-flowspec-mhalo_z0} show ISM outflow rates normalised by ISM gas mass (``specific outflow rates''). We find that specific outflow rates in satellite galaxies clearly increase with increasing halo mass -- which is not the case if existing ISM mass is not controlled for. This picture aligns with the expectation of increased ram-pressure and tidal stripping in group environments. We stress that we do not distinguish between different gas removal mechanisms -- but we remind the reader of the findings of \citet{Bahe2015}, who show that the majority of {\it excess} gas removal in satellites relative to field galaxies can be explained by ram pressure stripping, with a sub-dominant contribution from tidal stripping.

On average, outflow rates in satellite galaxies for all halo mass bins exceed those measured for central galaxies, with the exception of satellites with stellar mass \mstar$>11$. At $z\approx1$, central galaxies exhibit fairly constant specific outflow rates at $\dot{M}_{\rm out}/M_{\rm ISM}\approx0.4\ {\rm Gyr}^{-1}$ in galaxies below \mstar$\approx11$, effectively balancing specific replenishment rates. All satellite samples show steadily increasing specific outflow rates with halo mass -- with $\dot{M}_{\rm out}/M_{\rm ISM}$ values of $0.5\ {\rm Gyr}^{-1}$, $0.75\ {\rm Gyr}^{-1}$, and $1\ {\rm Gyr}^{-1}$ for the three halo mass samples from \mhalo$=11.5$ to \mhalo$=15$ respectively. For satellites in the large group/cluster-mass bin, this means that in the absence of star formation and inflows, such galaxies would have their full gas reservoir stripped within $1$ Gyr.

Adding the quantities in the $1^{\rm st}$ and $2^{\rm nd}$ rows of panels, we illustrate the total net rate of change in ISM mass of galaxies in the $3^{\rm rd}$ row of panels. This quantity  reflects the deviation of a galaxy from being in equilibrium -- with a net positive rate meaning that a galaxy would be expected to build up its ISM gas reservoir; and a net negative value meaning a galaxy would be net depleting its ISM reservoir. As per Equation \ref{eq:s3:tdep}, these net flow rates correspond to the inverse of the predicted depletion time-scale of the ISM, assuming that the flow rates remain constant. For all epochs shown, central galaxies generally show net flow values very close to equilibrium below \mstar$\approx11$. This is also true for satellite galaxies in the ~sub-$L^{\star}$ halo mass sample. In the two more massive halo mass samples, however, it is clear to see the combined influence of stripping and starvation on satellites -- particularly at $z\approx1$. 

Focusing on the small group-mass sample, \mhalo$\in[12.5,13.5)$ -- we record net $\dot{M}_{\rm ISM}/M_{\rm ISM}$ values of $\approx-0.6\ {\rm Gyr}^{-1}$ -- corresponding to an average depletion time of $\approx1.5$ Gyr (assuming the net specific flow rate remains constant); this value remaining relatively steady over stellar mass up to \mstar$\approx11$. If one were to assume that these satellites experienced the same specific replenishment as central galaxies; this would shift $\dot{M}_{\rm ISM}/M_{\rm ISM}$ values from $-0.6\ {\rm Gyr}^{-1}$ to $\approx-0.3-0.4\ {\rm Gyr}^{-1}$ -- corresponding to depletion times of $\approx2.5-3$ {\rm Gyr} on average. Thus, even though these galaxies do not experience sufficient suppression in gas accretion to quench via starvation alone (as per the positive specific replenishment rates), the moderate reduction in specific inflow rates can accelerate the depletion of ISM reservoirs by a factor of $\approx2$ on average. 

Focusing on the large group/cluster-mass sample, \mhalo$\in[13.5,15]$ at $z\approx1$ -- we note mild variation in $\dot{M}_{\rm ISM}/M_{\rm ISM}$ across stellar mass. Satellites at \mstar$=9$ and \mstar$=11$ see a net value of $\dot{M}_{\rm ISM}/M_{\rm ISM}\approx-1\ {\rm Gyr}^{-1}$, but between these mass scales, there is a slight peak in the net loss rate at \mstar$=10$, at $\dot{M}_{\rm ISM}/M_{\rm ISM}\approx-1.4\ {\rm Gyr}^{-1}$ (corresponding to a depletion time of $\approx0.7\ {\rm Gyr}$, assuming a constant specific outflow rate). This reflects the peak suppression of specific replenishment rates at this mass scale, where SFRs clearly exceed near-zero inflow rates. Assuming that such a satellite were to experience the same specific replenishment rates as centrals, the same net value of  $\dot{M}_{\rm ISM}/M_{\rm ISM}$ would shift to $\approx-0.8\ {\rm Gyr}^{-1}$ -- that is to say, the same rate of stripping would be capable of completely depleting the ISM of a central galaxy (in the absence of starvation) within $\approx1.2\ {\rm Gyr}$. Assuming a net balance between star formation and inflows and a constant specific outflow rate, outflows alone would deplete the ISM in $\approx1\ {\rm Gyr}^{-1}$. 

In general, we note that that there is notably increased $16{\rm th}-84^{\rm th}$ percentile spread in gas inflow, outflow, and net flow rates with increasing redshift. Given that these quantities are actually better sampled at higher redshift (due to systematically higher gas flow rates at early times), we argue that this increased variance is a true reflection of a galaxy population exhibiting real diversity in terms of evolutionary pathways. 

 In the most extreme examples of starvation --  \mstar$\approx10$ satellites in groups with \mhalo$>13.5$ at $z=1$ -- galaxies experience a median offset between inflow and SFR of $(\dot{M}_{\rm in}-\dot{M}_{\star})/M_{\rm ISM}\approx-0.4\ {\rm Gyr}^{-1}$. Assuming an outflow rate that matches the average central galaxy of the same mass, $\dot{M}_{\rm out}/M_{\rm ISM}\approx0.5\ {\rm Gyr}^{-1}$, the net (negative) specific flow rate does not exceed $\dot{M}_{\rm net}/M_{\rm ISM} = 1\ {\rm Gyr}^{-1}$. \textit{ This means that even in the most extreme cases of starvation, at least 50\% of satellites will still require enhanced outflows (i.e., direct cold gas stripping) to produce sub-Gyr indicative depletion time-scales}. Such depletion time-scales can be produced by stripping alone. However, this does not preclude the occurrence of starvation, which can moderately accelerate the process. 

\begin{table}
\centering
\includegraphics[width=1\columnwidth]{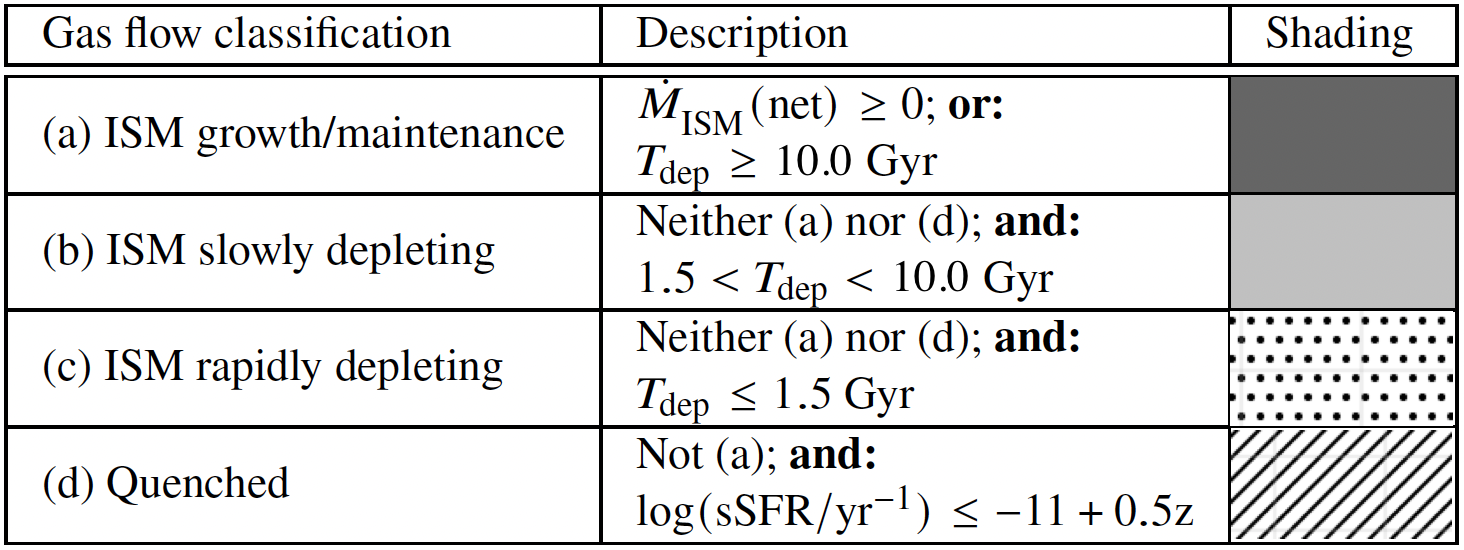}
\caption[Classification of galaxies into classes based on average gas flow rates.]{Different gas flow classifications based on average inflow, outflow, and SFRs. We class galaxies into those that are (a) net growing or maintaining their ISM reservoir, (b) slowly depleting their ISM reservoir, (c) rapidly depleting their ISM reservoir, and (d) quenched. The delineating depletion time-scale between ``slow'' and ''rapid'' quenching is 1.5 Gyr, based on the median quenching time-scale ($\tau_{\rm Q}$) value measured for \eagle\ galaxies in \citet{Wright2019}. The shading used for the histograms in Figs. \ref{fig:s3:mstar-emclass-mhalo_z} and \ref{fig:s4:trel-ormemclass_mstar} is provided in the right-hand column.} 
\label{tab:s3:emclass}
\end{table}

 With information about a galaxy's ISM mass, and its inflow, outflow and star formation rates (as presented in \ref{fig:s3:mstar-flowspec-mhalo_z0}) -- we can predict the mass of gas in the ISM of a galaxy at a time $t+\Delta t$ if we take Equation \ref{eq:isminst} and discretise in time:

\begin{equation}\label{eq:s3:ismdt}
M_{\rm ISM} (t+\Delta t) = M_{\rm ISM} (t) + [\dot{M}_{\rm in} (t) -\dot{M}_{\rm out} (t) -\dot{M}_{\star} (t)]\times \Delta t.
\end{equation}

If the quantity in the square brackets (inflow rate subtract outflow rate and star formation rate) is negative -- the ISM is being {\rm net} depleted, and we can calculate an indicative depletion time-scale by setting $M_{\rm ISM} (t+\Delta t)$ to $0$, solving for $\Delta t$:

\begin{equation}\label{eq:s3:tdep}
\Delta t \equiv T_{\rm dep}=\frac{M_{\rm ISM}} {\dot{M}_{\rm in}-\dot{M}_{\rm out}-\dot{M}_{\star}} = \frac{M_{\rm ISM}} {{\rm Net}\ \dot{M}_{\rm ISM}}; 
\end{equation}

{\noindent where a net flow rate of $\dot{M}_{\rm ISM}/M_{\rm ISM}=\pm 1$ corresponds to a galaxy in which the flow rate would add (remove) the present gas mass of the ISM in 1 Gyr. We stress that this parameterisation of $T_{\rm dep}$ corresponds to an indicative depletion time-scale, and makes the assumption that inflow, outflow, and star-formation rates remain constant. We also remark that we do not include the return of gas to the ISM via stellar winds in this calculation, however, have confirmed that this is a very minimal effect and seldom influences our conclusions. }

Building on the arguments presented above regarding net flow rates and depletion time-scales, it is possible to classify the evolutionary state of a galaxy into one of four categories, as outlined in Table \ref{tab:s3:emclass}. Galaxies can either be (a) net growing or maintaining their ISM reservoir, (b) slowly depleting their ISM reservoir, (c) rapidly depleting their ISM reservoir, or (d) quenched. Galaxies in which the inflow rate exceeds the sum of outflow rate and SFR are classified into (a), as are those that have a depletion time-scale greater than $10$ Gyr.\footnote{We choose to use an upper-limit of $10$ Gyr, as this corresponds to the maximum lookback time of our sample at $z\approx2$. Our results are qualitatively insensitive to this choice for values between $5$ and $13.7$ Gyr .} Galaxies with ISM depletion time-scales between $1.5-10$ Gyr are classified as ``slowly'' depleting, while those with depletion time-scales less than $1.5$ Gyr are classified as ``rapidly'' depleting. We use $1.5$ Gyr as the delineating time-scale based on the median quenching time-scale ($\tau_{\rm Q}$) value measured for \eagle\ satellite galaxies in \citet{Wright2019} defined by sSFR, though remark our results are qualitatively insensitive to this choice for values between $1$ and $4$ Gyr.

\begin{figure*}
\includegraphics[width=\textwidth]{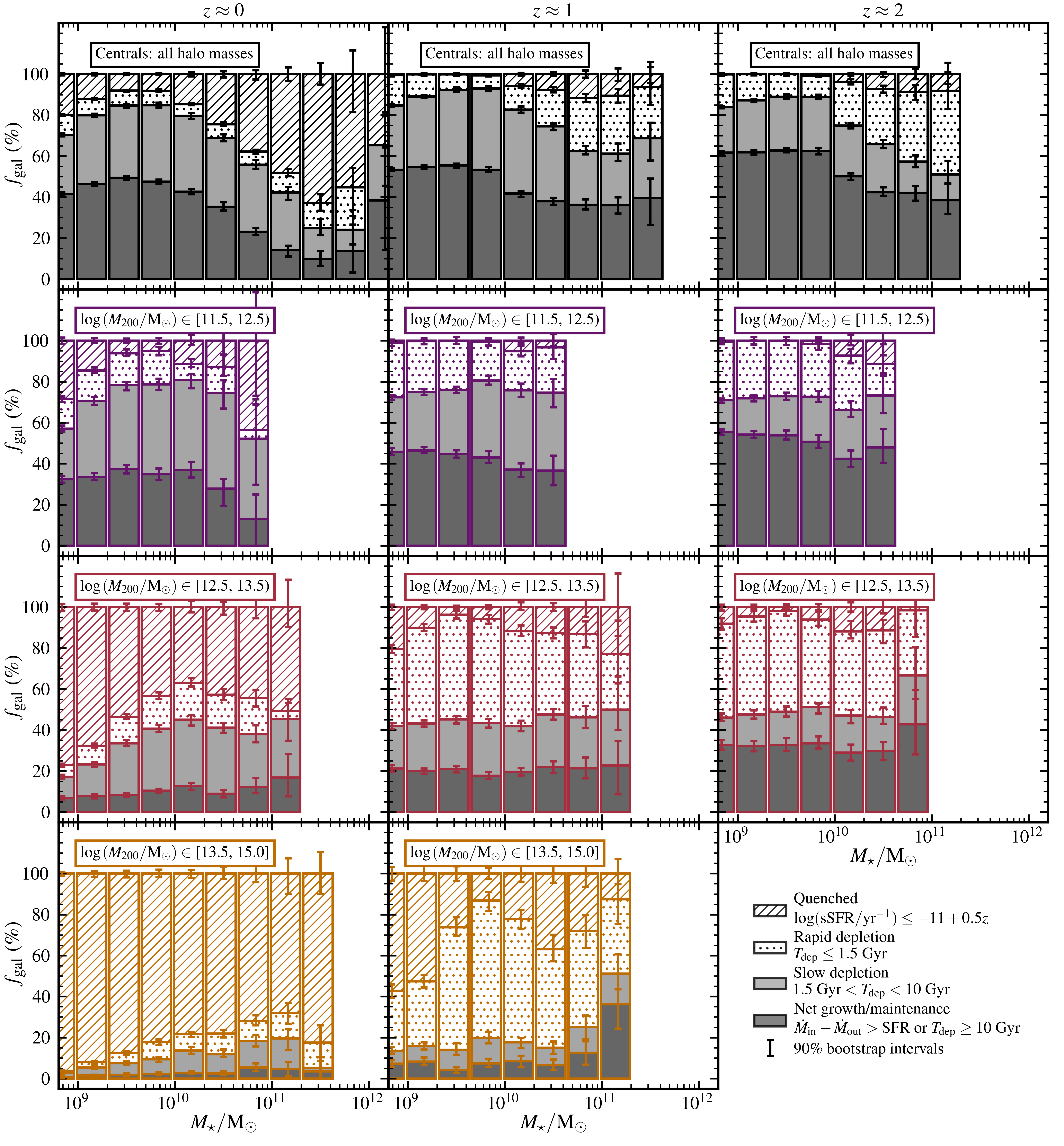}
\centering
\caption[Satellite and central gas flow status (Table \ref{tab:s3:emclass}) as a function of $M_{\star}$ in \eagle.]{The breakdown of galaxies into gas flow classes (outlined in Table \ref{tab:s3:emclass}), as a function of stellar mass (with 3 bins taken per decade in mass). A galaxy can be classified as either (a) net growing/maintaining its ISM reservoir -- dark grey shading; (b) slowly depleting its ISM reservoir -- light grey shading; (c) rapidly depleting its ISM reservoir -- dotted shading; and (d) quenched -- forward-hatched shading. The delineating time-scale between ``slow'' and ''rapid'' depletion is $1.5$ Gyr, based on the median quenching time-scale ($\tau_{\rm Q}$) value measured for \eagle\ galaxies in \citet{Wright2019}. Each column corresponds to a different epoch; including the $6$ snipshots closest to $z=0$, $z=1$ and $z=2$ from left to right. Each row corresponds to a different sample, with the top panels illustrating the breakdown for central galaxies, and the bottom $3$ rows showing the results for satellite galaxies in bins of increasing halo mass. We only include stellar mass bins where the statistics have been calculated with at least 20 galaxies. We do not include the results for satellites in the highest halo mass bin at $z\approx2$ due to low number statistics. Error-bars denote the bootstrap-generated $90\%$ confidence interval on the fractions (based on 500 re-samples). We exclude galaxies undergoing a major or minor merger from each of these samples.}
\label{fig:s3:mstar-emclass-mhalo_z}
\end{figure*}

As per the discussion in \S\ref{sec:introduction}, a galaxy experiencing reduced accretion rates and ``slow'' quenching would closely reflect the archetypal starvation scenario, compared to a galaxy experiencing ``fast'' quenching that closely reflects a ``stripping'' scenario. Our results above suggest that significant stripping is certainly required to produce $< {\rm Gyr}$ quenching time-scales, however does not preclude the suppression of gas accretion taking place. We also note that the indicative depletion time-scale of a galaxy based on instantaneous flow rates, in some cases, may systematically under-predict the time it takes to deplete a galaxy's ISM reservoir. This is because the rate of gas removal via both stripping and consumption in star formation are likely to scale down with the decreasing gas content of the ISM (and not remain at the initial high removal rates), and prolong the quenching phase. 

We classify galaxies as ``quenched'' (i.e., depleted) if they meet the simple criteria $\log({\rm sSFR}/{\rm yr}^{-1})<-11+0.5z$ (e.g. \citealt{Furlong2015}), and do not otherwise meet the criteria to be classified into group (a). Our choice to exclude ``rejuvenating'' galaxies from the quenched population means that group (d) does not exactly correspond to a classically defined ``quenched fraction'', however we remark that the proportion of rejuvenating galaxies in the samples considered is ubiquitously very low ($<1\%$). We also remark that our quenched definition, being based on sSFR, does not explicitly ensure that the ISM of a galaxy has been entirely depleted. Pragmatically, however, we have found that this cut seldom selects galaxies with gas fractions above $0.05$, and importantly allows us to determine the ``quenched'' status of galaxies with a commonly used observable property.

Using the arguments above, Fig. \ref{fig:s3:mstar-emclass-mhalo_z} shows the fraction of galaxies belonging to the gas flow classes outlined in Table \ref{tab:s3:emclass} as a function of stellar mass at $z\approx0$, $z\approx1$, and $z\approx2$. We split the sample of galaxies into the same selections illustrated in Figs. \ref{fig:s3:mstar-mgas-mhalo_xgass} and \ref{fig:s3:mstar-flowspec-mhalo_z0}; namely, centrals of all halo masses, and satellites in $3$ bins of halo mass between \mhalo$=11.5$ and $15$. 

Focusing first on central galaxies (top row), we note that the fraction of galaxies able to maintain or grow their ISM reservoir decreases with stellar mass. At $z\approx2$ the fraction of galaxies growing or maintaining their ISM sits at $60-65\%$ for galaxies with stellar mass below \mstar$=10$, with the ISM of approximately $20-30\%$ of galaxies classified as slowly depleting in the same mass range. For galaxies above \mstar$=10$ at $z\approx2$, the fraction drops towards $40\%$, with a much higher proportion of galaxies ($\approx 30\%$) experiencing rapid depletion. This is likely due to AGN feedback in this stellar mass range. Instead considering $z\approx0$ galaxies, $40-50\%$ of centrals are net growing their ISM reservoir below \mstar$\approx10$, and above this mass, the quenched fraction increases to $\approx60-70\%$. Aside from high-mass galaxies at $z\approx2$, it appears that there are ubiquitously more centrals in the slowly depleting group compared to the rapidly depleting group. This aligns with the quenching time-scales found in \citet{Wright2019}, who show that very few centrals quench in time-scales shorter than $\approx1$ Gyr. 

The behaviour of satellites in the lowest halo mass sample, \mhalo$\in[11.5,12.5)$ (second row), is very similar to the sample of central galaxies, with only very slight increases in the proportion of rapidly-quenching galaxies. This is true up to \mstar$\approx11$ at $z\approx0$, as very few satellites exist above this stellar mass in this sample. Galaxies in the next halo mass sample, \mhalo$\in[12.5,13.5)$ (third row), begin to show the clear signatures of environmentally-induced quenching. At $z\approx2$, just over $50\%$ of \mstar$<10.5$ satellite galaxies in this sample exist in the  rapidly depleting group -- much higher than the slowly depleting fraction at $\approx 10-20\%$. The quenched fraction grows correspondingly towards $z\approx0$, with the fraction of satellites quenched largest ($>50\%$) for \mstar$<9.5$ (due to environmental effects) and \mstar$>11$ (due to AGN feedback).

Galaxies in the highest halo mass bin, \mhalo$\in[13.5,15]$ (bottom row), corresponding to larger groups and clusters, show even more dramatic signatures of environmental quenching, with $\approx80\%$ of the non-quenched sample at \mstar$\approx10$ experiencing rapid ISM depletion. Very few satellites in this sample are able to net grow or maintain their ISM reservoir: $\approx5\%$ at $z\approx1$, and even less at $z\approx0$. At $z\approx0$, we measure that approximately $80\%$ of satellites are quenched (averaged across stellar mass), with the fraction highest for galaxies with \mstar$<9.5$. As we show in \S\ref{sec:results:orbits}, the small fraction of un-quenched satellites in this sample is likely a reflection of the population of satellites that have not yet entered the host's virial radius.


\section{Gas flows along the infall and orbit of satellite galaxies}\label{sec:results:orbits}

In this section, we move from the static analysis of \S\ref{sec:results:mhalo} to a temporal analysis of the gas mass, flow rates and star formation rates in satellite galaxies along their infall and orbits in group and cluster environments. We again stress that while previous work has separately explored inflow, outflow and star formation rates in satellite galaxies, we aim to holistically consider these processes and their combined role in regulating the gas content of satellite galaxies. 

To identify the infall and orbits of satellites in \eagle, we make use of the \orbweaver\ tool as described in \S\ref{sec:methods:orbweaver} \citep{Poulton2019,Poulton2020}. In the case where multiple orbital hosts for a single galaxy have been identified, we only consider the most massive host. We then select galaxies for our orbital analysis by applying the following criteria {\it at the first pericentric passage (fp)} for each galaxy (and quote the percentage of fps for which this criterion is met):

\begin{itemize}
    \item $z_{\rm fp}^{}\leq2$ ($93.9 \%$) -- ensures that particle data is available;
    \item Host \mhalo$\geq12.5$ ($54.2 \%$) -- expect significant environmental effects;
    \item Satellite \mstar$\in[8,11]$ ($94.6\%$) -- removes very small and very massive satellites;
    \item  Orbital radius within $R_{\rm 200}$ of host ($57.2\%$) -- consider only close approaches.
\end{itemize}

This extracts a total of $7\ 803$ first pericentric passages in the \eagle\ L100-REF box. We do not include satellites with a (pericentric) stellar mass above $10^{11}{\rm M}_{\odot}$ as there are very few, and in these circumstances the central-satellite delineation can be confused between outputs (e.g. \citealt{Canas2019}). For the purposes of our analysis, we also require that each satellite reaches at least one apocenter post-infall, and that it is not lost in the merger trees prior to this point -- reducing the sample to $2\ 810$ orbits. This reduction in sample size is due to several factors. Firstly, some satellites are disrupted  prior to apocenter (i.e. destroyed, or merged with the host central). Secondly, given the high-density environment and interactions between haloes, the merger trees may struggle to match haloes across snapshots. Thirdly, many ($\approx2\ 000$) of the orbits removed were recent infalls, with pericenters at lookback times of $\leq1$ Gyr -- meaning these galaxies were not necessarily expected to reach an apocenter by $z=0$. We do not include these late-infall galaxies in our sample, meaning our analysis is slightly biased to higher-redshift infall events (as explicitly shown in Fig. \ref{fig:s4:mstar-orbprops}). 

Lastly, we remove the satellites that have been ``pre-processed'' (i.e., have been a satellite in a host halo with $M_{200}>10^{12}{\rm M}_{\odot}$) prior to entering the current orbital host. This allows us to analyse how environment affects satellites on their {\it first-infall} to a group environment,  and reduces our final sample to $2\ 054$ orbits. This pre-processed fraction, $\approx30\%$, is slightly lower than that found previously by \citet{Bahe2019} using the Hydrangea simulations, with fractions typically above $50\%$ (aside from higher mass satellites in smaller groups). We note, however, that we only consider pre-processing to have occurred if the previous host was above a mass of $10^{12}{\rm M}_{\odot}$. Below this halo mass, based on the results presented in \S\ref{sec:results:mhalo} and Fig. \ref{fig:s3:mstar-emclass-mhalo_z} (considering the lowest halo mass sample), we argue that the influence of environment is small due to the similarity between the properties of this sample of satellites and the sample of central galaxies. In addition, Hydrangea corresponds to cluster zooms and as such are biased to high density environments, where more pre-processing is expected, compared to the \eagle\ box.

With the orbital characteristics of our sample generated with \orbweaver, we are able measure an orbital time-scale and thus meaningfully analyse how the properties of the satellite change as a function of orbital time and phase. 

\subsection[Individual examples of satellite infall to groups in \eagle]{Individual examples of satellite infall to  groups in \eagle}\label{sec:s4:vis}

\begin{figure*}
\includegraphics[width=0.98\textwidth]{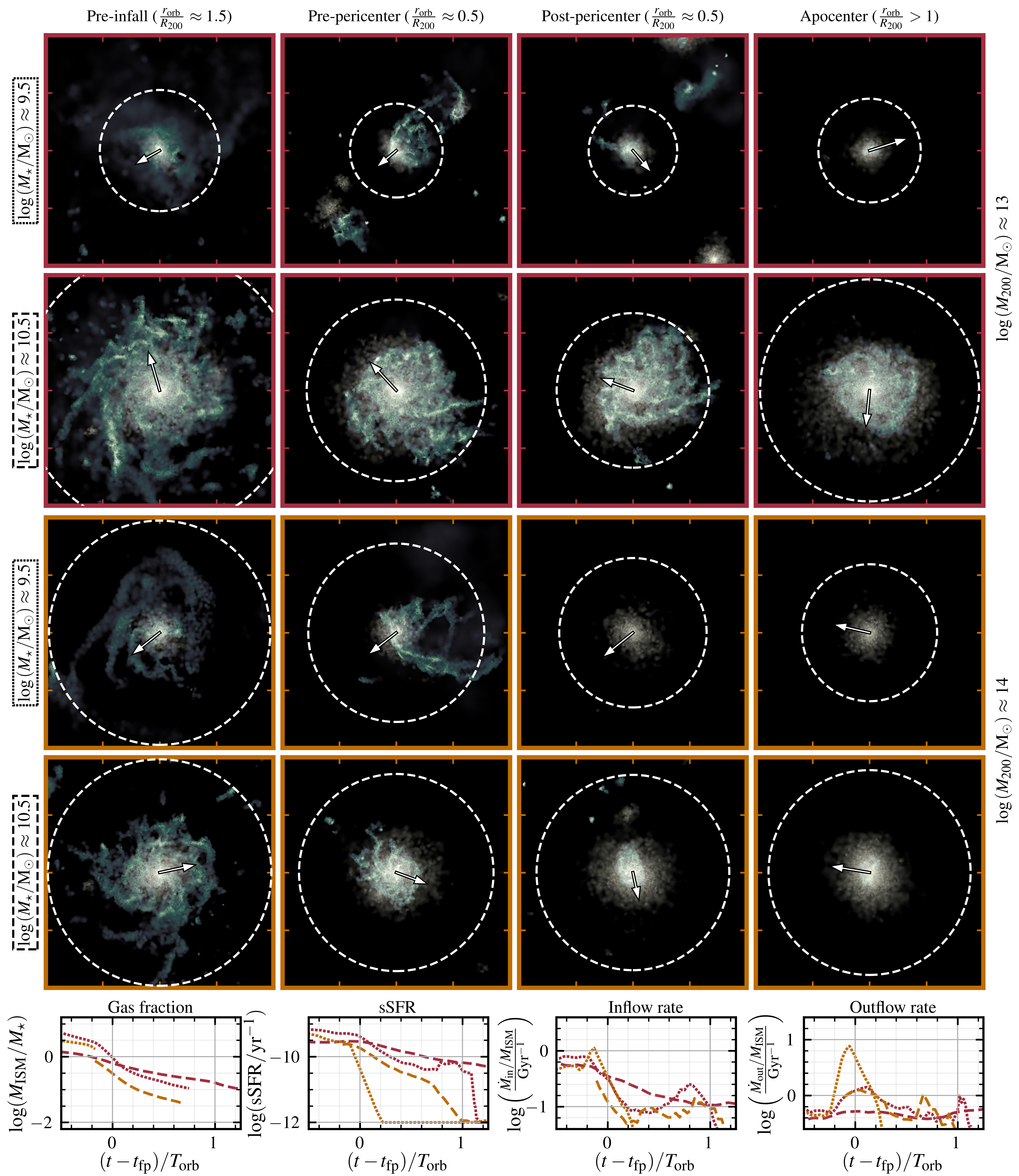}
\caption[Example visualisations of satellites on their infall to group and cluster-mass haloes.]{Example visualisations of satellites on their infall to group and cluster-mass haloes. Each row corresponds to the evolution of a single galaxy, tracked through the merger trees along its orbit. The top two rows of visualisation panels show example galaxies falling onto groups with mass \mhalo$\in[12.5,13.5)$, and the bottom two rows of visualisation panels show galaxies falling into groups with mass \mhalo$\in[13.5,15]$. The first and third rows of panels correspond to example galaxies at \mstar$\approx9.5$ (dotted lines in bottom panels) at first pericenter; while the second and fourth rows correspond to example galaxies at \mstar$\approx10.5$ (dashed lines in bottom panels). We remark that the stellar mass of these galaxies do not markedly change post-infall. From left to right, we include a visualisation of the satellite (i) pre-infall at $r_{\rm orb}/R_{200}=1.5$, (ii) pre-pericenter $r_{\rm orb}/R_{200}=0.5$, (iii) post-pericenter $r_{\rm orb}/R_{200}=0.5$, and (iv) at first apocenter. Axes ticks correspond to increments of $20$ pkpc. Stellar mass density and gas mass density are shown in pale yellow and green respectively, using the same logarithmic scaling. The velocity vector of the satellite's center of mass and its relative magnitude in the image plane are included at the center of each panel; and the effective baryonic radii of the galaxies, $R_{\rm BMP}$, is illustrated with dashed lines. The bottom row of panels illustrates different properties of the satellites along their orbit, namely (from left to right): gas fraction, specific star formation rate, ISM inflow rate, and ISM outflow rate.}
\label{fig:s4:example-infall_mstar}
\end{figure*}

For illustrative purposes, we include example visualisations of the infall of a number of satellites part of our sample in Fig. \ref{fig:s4:example-infall_mstar}. Each row of visualisation panels shows the same satellite along its orbit, from left to right: pre-infall ($\approx1.5R_{200}$),  pre-pericenter ($\approx0.5R_{200}$), post-pericenter ($\approx0.5R_{200}$), and at first apocenter ($> R_{200}$). The $4$ panels at the bottom of Fig. \ref{fig:s4:example-infall_mstar} illustrate the gas fractions, sSFRs, inflow rates, and outflow rates of each of the satellites shown. 

We first focus on the top row of panels -- an $M_{\star}\approx10^{9.5}{\rm M}_{\odot}$ satellite on infall to a $M_{200}\approx10^{13}{\rm M}_{\odot}$ halo. Pre-infall, the galaxy has a gas fraction (gas mass normalised by total baryonic mass) of $\approx0.8$, an ${\rm sSFR}$ of $\approx10^{-9}\ {\rm yr}^{-1}$, and its ISM inflow rate exceeds the outflow rate. Prior to pericenter, the radial extend of the galaxy's gaseous component appears to exceed that of the stellar component. Once the galaxy enters $R_{\rm 200}$ in the second panel, there is evidence of a clearly disturbed gas morphology and a ram-pressure stripping tail opposite to the infall velocity vector. By this point, the galaxy's gas fraction has begun to drop steeply; however the ${\rm sSFR}$ appears to only gradually reduce, even seeing a slight peak prior to pericenter as the gaseous leading edge of the galaxy is compressed by interaction with the intra-group medium. This agrees with the findings of \citet{Troncoso2020}, who found using the \eagle\ simulations that  the leading half of satellite galaxies (as defined by the trajectory of the galaxy in the halo) in high mass groups has a higher ISM pressure than the trailing half -- consistent with this observed compression.

Post-pericenter, the galaxy's gas fraction visibly drops (to $\approx0.25$), and the ${\rm sSFR}$ begins to show a steeper decline. Inflow rates drop by $\approx1$ dex, while the outflow rate peaks close to pericenter, $\approx0.5$ dex above the pre-infall value. At first apocenter, there is very little gas remaining in the galaxy (though not zero, as per the gas fraction panel), and the  ${\rm sSFR}$ continues to steadily drop. Outflow rates decrease back to just above pre-infall values, however inflow rates remain very low -- $\approx1$ dex below pre-infall. After one pericentric passage, we note that the radial extent of the gaseous disk has contracted to roughly match the extent of the stellar disk. 

The second row of panels shows a more massive $M_{\star}\approx10^{10.5}{\rm M}_{\odot}$ galaxy on infall to a similar mass, $M_{200}\approx10^{13}{\rm M}_{\odot}$ halo. Pre-infall (at $\approx1.5R_{\rm 200}$), the galaxy has a gas fraction and sSFR slightly lower than its lower mass counterpart at $\approx0.6$, consistent with the global mass -- sSFR relation. When the galaxy enters $R_{\rm 200}$, we can see the gaseous component of the galaxy becomes slightly compressed on the leading edge of the disk, and slightly extended in the trailing direction. The rate of sSFR decline remains relatively steady on infall though, and outflow rates show a very slight increase around pericenter, $\approx0.1-0.2$ dex above pre-infall values. After pericenter, we note a steeper decrease in galaxy gas fraction, and that the reduction in ${\rm sSFR}$ begins to accelerate again. Inflow rates steadily drop after pericenter to $\approx0.5$ dex below pre-infall values at apocenter, where the galaxy's gas fraction has dropped to $\approx0.3$. While environment has a clear influence on the quenching of this galaxy, the decline in sSFR and gas fraction is relatively gentle, driven primarily by consumption of existing gas (starvation) after entry to the halo. Helping this gentle decrease in SFR is the lack of strongly enhanced outflow rates after pericenter. 

The third row of panels illustrates an  $M_{\star}\approx10^{9.5}{\rm M}_{\odot}$ satellite on infall to a more massive, $M_{200}\approx10^{14}{\rm M}_{\odot}$ halo. Pre-infall, the galaxy has a gas fraction and sSFR ($\approx0.8$ and $\approx10^{-9}\ {\rm Gyr}^{-1}$ respectively) similar to the galaxy of roughly equal mass in the top row of panels. Additionally, the inflow rate at $(t-t_{\rm fp})/T_{\rm orb}\approx-1$ of $\approx10^{0}\ {\rm Gyr}^{-1}$ clearly exceeds the outflow rate, $\approx10^{-0.3}\ {\rm Gyr}^{-1}$. In the first visualisation panel, the galaxy shows little visual sign of disturbance on its infall. In the second panel, pre-pericenter, we see a very clearly disturbed gas morphology, with gas heavily compressed on the leading edge of the galaxy and extended in the trailing direction. This is reflected in massively increased outflow rates, peaking at $\approx10^{1}\ {\rm Gyr}^{-1}$ at pericenter -- which would correspond to complete removal of the cold gas reservoir within $100\ {\rm Myr}$, even in the absence of star formation. In the third panel, we can clearly see that there is no gas remaining -- with sSFR and gas fractions dropping near to zero very shortly after pericentric passage. After first pericenter, we also note that inflow rates to the ISM remain below $\approx10^{-1}\ {\rm Gyr}^{-1}$, meaning there is very little change for rejuvenation. 

Lastly, the fourth row of panels illustrates an $M_{\star}\approx10^{10.5}{\rm M}_{\odot}$ satellite on infall to an $M_{200}\approx10^{14}{\rm M}_{\odot}$ halo. In the first panel, we observe a clear disk-like morphology, with the galaxy's gaseous component extending radially further than the more compact stellar component. The galaxy shows a similar pre-infall gas fraction and sSFR  ($\approx0.6$ and $\approx10^{-9.5}\ {\rm Gyr}^{-1}$ respectively) to its equal-mass counterpart in the second row of panels. In this large host, however, this gas fraction and sSFR actually begin to drop even prior to pericenter -- as reflected in the second panel where the outer gas disk has been stripped to roughly match the radial extent of the stellar component. Post-pericenter, in the third panel, we observe even more gas removal -- with a clearly disturbed gas morphology embedded in the mostly undisturbed stellar disk.  While a large amount of gas has been removed, the gas content of the galaxy does not quite drop to the lower limit. At apocenter, the galaxy retains a gas fraction of $<0.1$ and a low (but non-zero) sSFR of $\approx10^{-10.5}\ {\rm yr}^{-1}$. In this case, the deeper potential of the stellar reservoir offers a higher restoring force against stripping.

\subsection[Statistical analysis of satellite infall to groups in \eagle]{Statistical analysis of satellite infall to groups in \eagle}\label{sec:s4:stat}

\begin{figure}
\includegraphics[width=1\columnwidth]{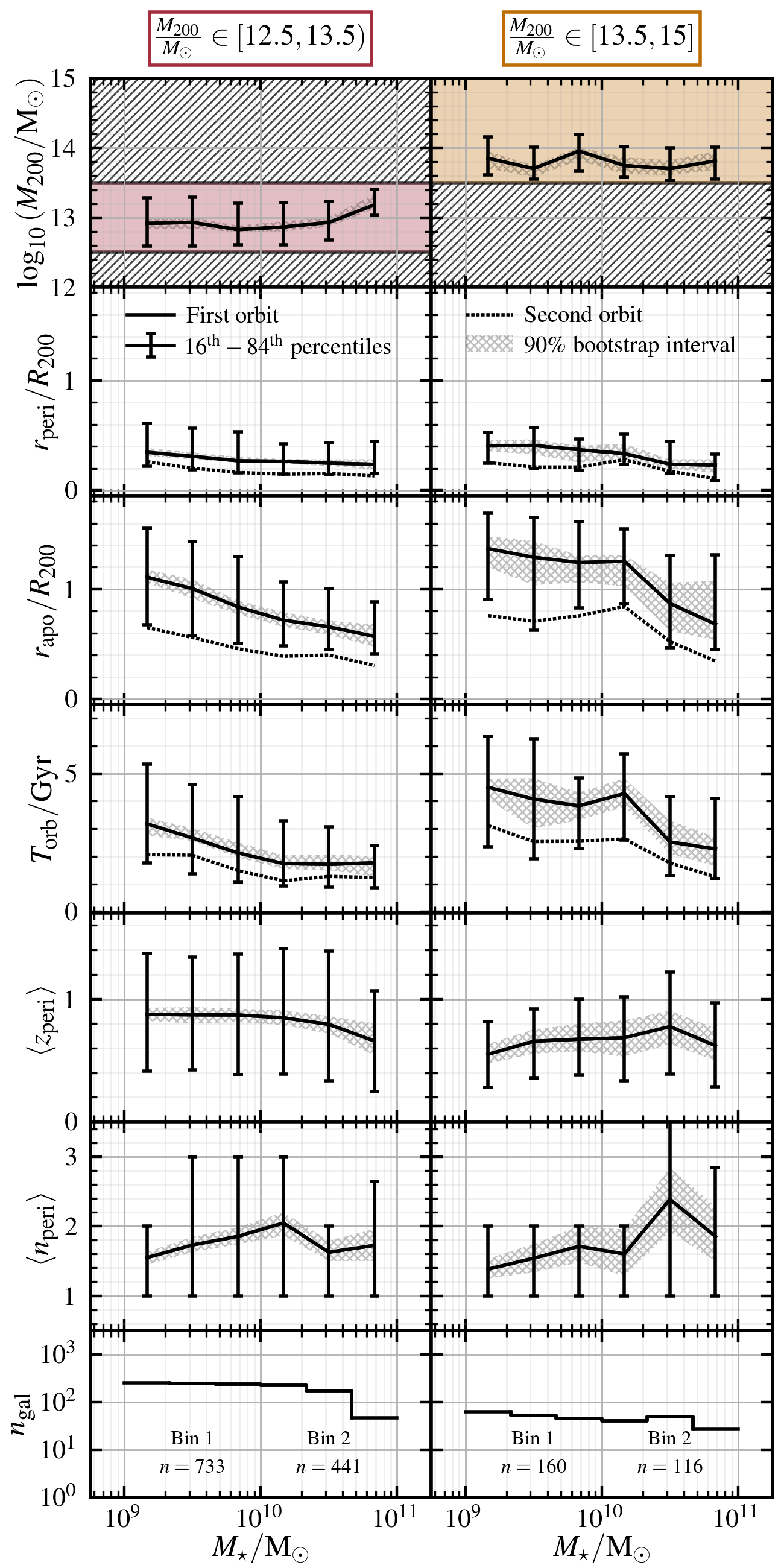}
\caption[Properties of the selected infall orbits as a function of stellar mass.]{Properties of the selected infall orbits as a function of stellar mass. Each column pertains to different bins in host mass -- left including satellites with host masses $M_{\rm 200}/{\rm M}_{\odot}\in [10^{12.5},10^{13.5})$ -- red; and right including host masses $M_{\rm 200}/{\rm M}_{\odot}\in [10^{13.5},10^{15}]$ -- orange. From top to bottom, each row illustrates: host halo mass, $M_{\rm 200}$);  pericentric distance normalised by host virial radius, $r_{\rm peri}/R_{\rm 200}$; apocentric distance normalised by host virial radius, $r_{\rm apo}/R_{\rm 200}$; orbital time-scale, $T_{\rm orb}$; mean redshift of first pericenter, $\langle z_{\rm peri} \rangle$; mean number of pericenters survived (before halo lost or merged with central), $\langle n_{\rm peri}\rangle$; and a histogram of the number of satellites included in the sample as a function of stellar mass. Error-bars display the $16^{\rm th}-84^{\rm th}$ percentile range in each stellar mass bin, and cross-hatched regions denote the bootstrap-generated $90\%$ confidence interval on the medians (based on 500 re-samples). We only include bins where the statistics have been calculated with at least 20 galaxies.}
\label{fig:s4:mstar-orbprops}
\end{figure}

In this subsection, we now move to statistically analyse our sample of satellite orbits. The orbital properties of the final sample as a function of stellar mass for each halo mass selection are shown in Fig. \ref{fig:s4:mstar-orbprops}. We illustrate these orbital properties to demonstrate the potential biases introduced in each sample in the process of binning the orbits by pericentric stellar mass (as is done for the remainder of this section). The left-hand panel shows the results for the small group sample (red), and the right-hand columns provides the results for the larger group/cluster selection (orange). The number of satellites included for each stellar mass bin in the two halo mass samples is quoted in the bottom row of panels; with there being $\approx3$ times as many orbits recorded for the group-mass sample compared to the larger group/cluster-mass sample. 

We remark that our initial orbit selection included galaxies with stellar mass between $10^{8}{\rm M}_{\odot}$ and $10^{9}{\rm M}_{\odot}$ -- however, we choose to focus on galaxies with $M_{\star}\geq10^{9}{\rm M}_{\odot}$ such that there are ample gas and stellar particles in the galaxies and the influence of numerical effects is minimised.

The top row of panels in Fig. \ref{fig:s4:mstar-orbprops} illustrates that the two aforementioned halo mass selections do not introduce a significant bias in halo mass as a function of stellar mass. The group-mass sample shows a steady median halo mass at \mhalo$\approx13$, with a very slight upturn at \mstar$\approx11$. This slight upturn is likely due to the low probability of a halo with mass on the lower end of the $10^{12.5}-10^{13.5}{\rm M}_{\odot}$ sample hosting a massive satellite, with stellar mass in excess of $\approx10^{10.5}{\rm M}_{\odot}$. The larger group/cluster sample (orange, right panel) also shows a steady median across stellar mass, slightly below \mhalo$\approx14$.

The second row of panels in Fig. \ref{fig:s4:mstar-orbprops} shows the first pericentric radius of satellite orbits in the two halo mass samples as a function of satellite stellar mass. For those satellites that survive to a second pericenter, we illustrate the mean second pericentric distance with dashed lines. In both halo mass samples, we observe that the measured pericentric radii decrease with increasing stellar mass -- from $r_{\rm peri}/R_{200}\approx0.5$ at \mstar$\approx9$ to $r_{\rm peri}/R_{200}\approx0.25$ at \mstar$\approx11$. This bias is likely introduced by dynamical friction, as well as our requirement that the satellite survives pericenter and experiences at least one apocenter.

At smaller pericentric radii, the likelihood for satellite disruption is increased, particularly below $r_{\rm peri}/R_{200}\approx0.3$ -- roughly corresponding to the expected BMP radius of a central galaxy with \mstar$>10-11$ (see bottom panel of Fig. \ref{fig:s2:bmpstats}). More massive satellites -- with deeper gravitational wells and larger restoring forces -- are less likely to be entirely disrupted by such passages, however this is not the case for low-mass satellites with \mstar$\lesssim9.5$ (see Figure 6, dark green curve in \citealt{Bahe2019}). This may also be a result of numerical issues given the small number of stellar and gas particles in these galaxies ($<1000$); and the fact that such systems are more likely to be lost in merger trees. We note that second pericentric distances are systematically lower than the original pericenter -- decreased by a relatively steady fractional change in $r/R_{\rm 200}$ of $\approx 0.2$ for both halo mass bins, likely due to the effects of dynamical friction. 

The third row of panels in  Fig. \ref{fig:s4:mstar-orbprops} illustrates the first apocentric radius of satellite orbits. For those satellites that survive to a second apocenter, we also illustrate the mean second apocentric distance with the dashed lines. Similar to the pericentric distances discussed, there is a preference for low-mass galaxies to have larger apocentric distances -- in this case extending to beyond the host virial radius for satellites with stellar mass below $10^{9}{\rm M}_{\odot}$. The difference between first and second apocentric distance is pronounced, with many second apocenters halving their halo-centric distance after completing a full orbit. 

The larger orbital radius noted for low-mass satellites in rows 2 and 3 of Fig. \ref{fig:s4:mstar-orbprops} translates to longer orbital time-scales for these galaxies, which is explicitly demonstrated in the fourth row of panels. We measure orbital time-scales in our sample by doubling the time between pericenter and apocenter in the first orbit; and use the same procedure for the galaxies that survive a second orbit. We note that the precision of these measurements is limited by the cadence of the \eagle\ snipshots -- ranging between $50$ and $150$ Myr. For both halo mass samples, orbital time-scales decrease from $T_{\rm orb }\approx4$ Gyr at \mstar$\approx9$ to $T_{\rm orb}\approx2$ Gyr at \mstar$\approx11$. For galaxies that survive a second orbit, their orbital time-scales, on average, decrease by $\approx1$ Gyr across stellar mass for the group-mass sample (left), and slightly more for the larger group/cluster sample (right) where the effects of dynamical friction are more significant. We note that it is important to consider this bias for longer orbital time-scales in the lower-mass satellites when interpreting the results in the remainder of this section. 

The fifth row of panels in Fig. \ref{fig:s4:mstar-orbprops} shows the mean redshift of first pericenter as a function of stellar mass. For both samples we remark that there is no obvious bias (within the uncertainty) in accretion times of the satellites -- with the mean redshift steady at $z\approx 0.7-0.8$ for the smaller group-mass sample and $z\approx 0.6$ for the larger group/cluster-mass sample. The lower pericentric redshift recorded for the cluster-mass sample is a reflection of the assembly age being anti-correlated with halo mass \citep{DeLucia2007}. We also note that the $16^{\rm th}-84^{\rm th}$ percentile spread in pericentric redshifts for the smaller group-mass sample is much higher than the cluster-mass sample, particularly towards higher redshift ($z\approx1.5$). 

\begin{figure*}
\includegraphics[width=0.98\textwidth]{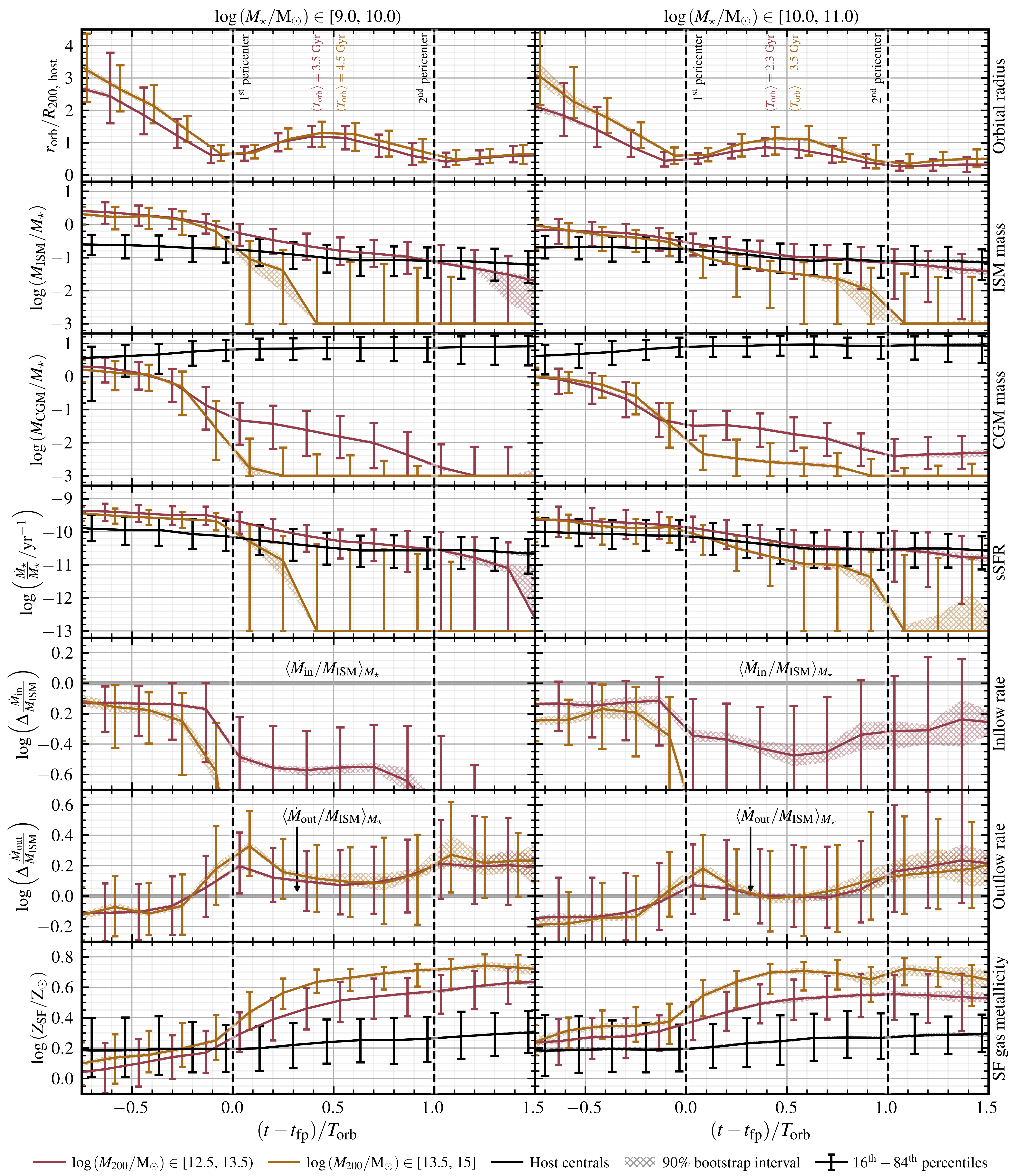}
\caption[The properties of \eagle\ satellites as a function of orbital time.]{Various properties of satellite galaxies and host centrals as a function of orbital time, relative to first pericenter. In each panel, the results are shown for satellites with host masses ($M_{\rm 200}$) between $10^{12.5}{\rm M}_{\odot} - 10^{13.5}$ ${\rm M}_{\odot}$ in red; $10^{13.5}{\rm M}_{\odot} - 10^{15}{\rm M}_{\odot}$ in orange; and the median values for the central galaxies of these satellites (combined from both halo mass samples) in black. Each column shows the results for two bins in satellite stellar mass (at pericenter) -- from left to right, \mstar$\in [9,10)$; and \mstar$\in [10,11)$. We stress that the selection of the host centrals for each sample is based on group mass alone, and that for the halo mass range considered $(M_{\rm 200}\geq10^{12.5})$, such centrals will typically possess stellar mass of order $M_{\star}>10^{10.5}{\rm M}_{\odot}$. From top to bottom, each row illustrates the following median values along the orbit: orbital radius relative to host virial radius, $r_{\rm orb}/R_{200}$; ISM gas fraction, $M_{\rm ISM}/M_{\star}$; CGM mass (normalised by stellar mass), $M_{\rm CGM}/M_{\star}$; ${\rm sSFR}=\dot{M}_{\star}/M_{\star}$; {\it excess} ISM inflow rate, $\Delta \dot{M}_{\rm ISM,\ in}/M_{\rm ISM}$; {\it excess} ISM outflow rate, $\Delta \dot{M}_{\rm ISM,\ out}/M_{\rm ISM}$; and gas-phase metallicity of star-forming particles, $Z_{\rm SF}/{\rm Z}_{\odot}$. Error-bars display the $16^{\rm th}-84^{\rm th}$ percentile range in each stellar mass bin, and cross-hatched regions denote the bootstrap-generated $90\%$ confidence interval on the medians (based on 500 re-samples). We only include bins where the statistics have been calculated with at least 20 galaxies.}
\label{fig:s4:trel-orbprops}
\end{figure*}

\begin{figure*}
\includegraphics[width=1.0\textwidth]{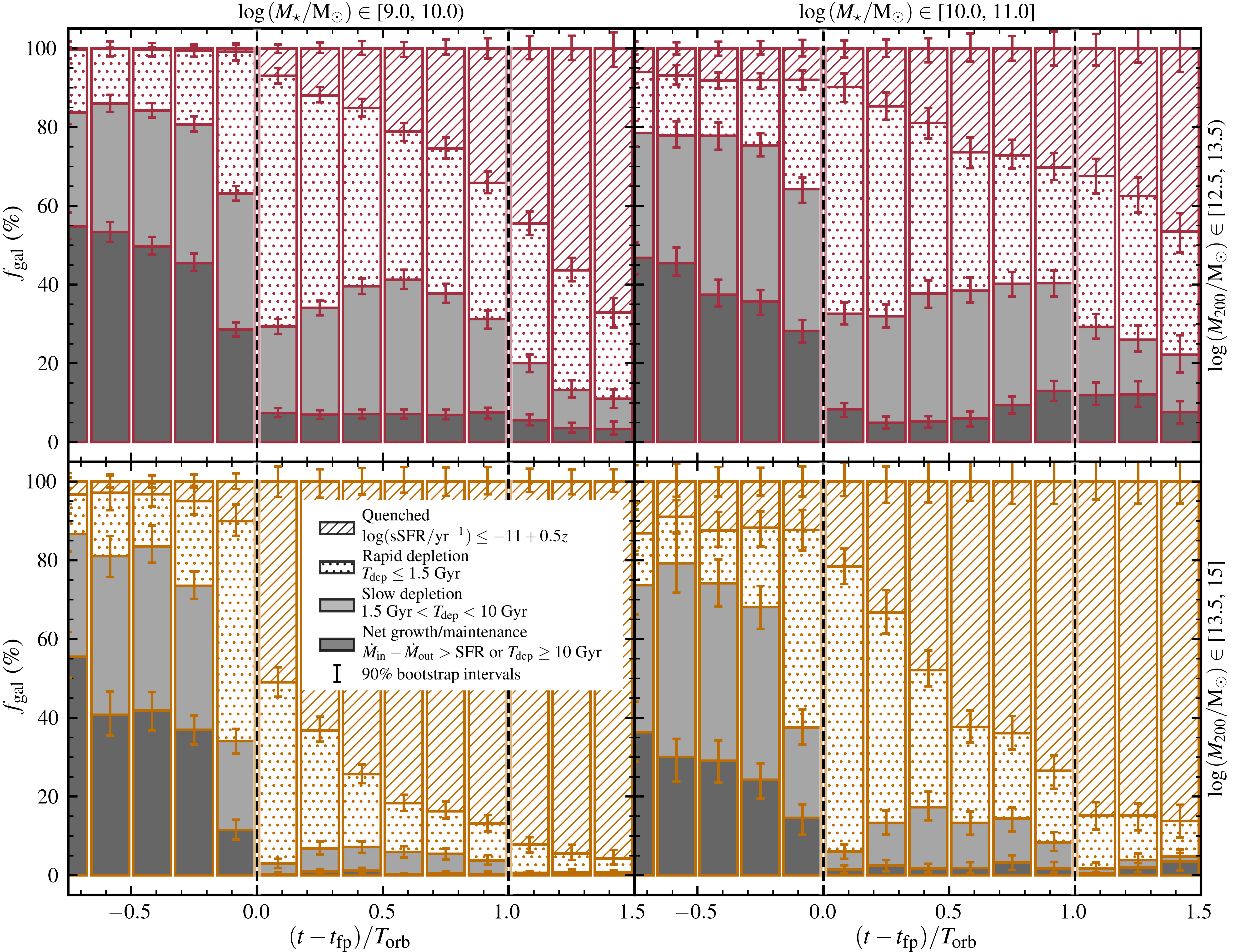}
\caption[Satellite gas flow status (Table \ref{tab:s3:emclass}) as a function of orbital time in \eagle.]{The breakdown of galaxies into gas flow classes (outlined in Table \ref{tab:s3:emclass}) as a function of orbital time relative to first pericentric passage. A galaxy can be classified as either (1) net growing/maintaining its ISM reservoir -- dark grey shading; (2) slowly depleting its ISM reservoir -- light grey shading; (3) rapidly depleting their ISM reservoir -- dotted shading; and (4) quenched -- forward-hatched shading. The delineating time-scale between ``slow'' and ''rapid'' depletion is $1.5$ Gyr, based on the median quenching time-scale ($\tau_{\rm Q}$) value measured for \eagle\ galaxies in \citet{Wright2019}. Each column corresponds to a different bin in stellar mass, from left to right \mstar$\in[9,10)$; and \mstar$\in[10,11]$. Each row corresponds to a different halo mass bin, with the top panels illustrating the breakdown for satellites in group-mass haloes, \mhalo$\in[12.5,13.5)$, red; and the bottom panels for satellites in larger group/cluster-mass haloes, \mhalo$\in[13.5,15]$. Error-bars denote the bootstrap-generated $90\%$ confidence interval on the fractions (based on 500 re-samples).}
\label{fig:s4:trel-ormemclass_mstar}
\end{figure*}

The sixth row of panels in Fig. \ref{fig:s4:mstar-orbprops} shows the mean number of pericenters survived as a function of stellar mass for each halo mass sample. Our selection requires at least one pericenter and apocenter to be survived, and thus $1$ becomes the lowest possible $n_{\rm peri}$ value in our sample. In both samples, we unsurprisingly observe that the number of pericenters survived before disruption (as per the \subfind\ trees) increases with stellar mass, from on average $\approx1$ passage at \mstar$\approx9$ to $\approx2$ passages at \mstar$\approx11$. For \mstar$>10.5$, the upper percentile can extend to $3-4$ pericentric passages. This simply means that as we move towards larger time since pericentric passage (as in Figs. \ref{fig:s4:trel-orbprops} and \ref{fig:s4:trel-ormemclass_mstar}), our sample size, and thus statistical power, begins to decrease.

Finally, the bottom row of panels in Fig. \ref{fig:s4:mstar-orbprops} show the number of satellites included in the sample as a function of stellar mass (again using 3 bins per decade in stellar mass). In the smaller group-mass sample, the number of galaxies is largest in the stellar mass range \mstar$\in[9,10]$ which contains $735$ satellites -- however the peak is not dramatic, with the surrounding bins still containing $\approx450$ satellites. In the larger group/cluster-mass sample, the number of galaxies is comparatively independent of stellar mass (with $\approx120-160$ satellites included in the sample per decade in stellar mass). 

With our selection in place, we move on to conduct a statistical analysis of satellite orbits by breaking our sample into $2$ bins of stellar mass -- (i) intermediate-mass satellites, \mstar$\in[9, 10)$; and (ii) high-mass satellites, \mstar$\in[10, 11]$. We remind the reader that our initial orbit selection included galaxies with stellar mass between $10^{8}{\rm M}_{\odot}$ and $10^{9}{\rm M}_{\odot}$ -- however, we choose to focus on galaxies with $M_{\star}\geq10^{9}{\rm M}_{\odot}$ to reduce the influence of any numerical effects. In addition to stellar mass, we delineate between satellites orbiting hosts in small group haloes, \mhalo$\in [12.5,13.5)$ -- as per the second halo mass bin/red sample in \S\ref{sec:results:mhalo}; and larger group/cluster-mass haloes with \mhalo$\in [13.5,15]$ -- as per the third halo mass bin/orange sample in \S\ref{sec:results:mhalo}. 

Moving forwards, splitting our selection into the aforementioned stellar and halo mass samples, we display a number of satellite properties as a function of orbital time in Fig. \ref{fig:s4:trel-orbprops}. In line with the work of \citet{Oman2021}, we use the first pericentric passage as an objective time ``zero point''  -- and also normalised by orbital time-scale to define the x-axis: $(t-t_{\rm fp})/T_{\rm orb}$. As outlined above, we calculate the orbital time-scale of satellites by doubling the interval between pericenter and apocenter for each orbit. Since we remove satellites that have been pre-processed, we note that in each satellite sample, the values presented in each row of panels at $(t-t_{\rm fp})/T_{\rm orb}\approx-0.75$ (left-most bin) approximately represent what would be expected of a field central galaxy with the same stellar mass (and are thus similar for both host mass samples pre-infall). We also remark that these pre-infall galaxies may indeed be ostensibly classified as central, with the requirement for selection in the satellite sample only enforced on satellite/central status at first pericenter. Based on the top row of panels, orbital distances pre-infall are, on average, outside $2-3 R_{\rm 200}$ -- where environmental effects are expected to be small (except when considering the hot gas reservoirs of lower mass satellites, \citealt{Bahe2013}). We note that the top row of panels also quotes the average orbital period for the first and second orbits of each sample.

We remind the reader that for inclusion in our sample, we require that a satellite reaches at least one apocenter in the merger trees -- i.e. to $(t-t_{\rm fp})/T_{\rm orb}=0.5$. For $(t-t_{\rm fp})/T_{\rm orb}>0.5$, we include the results for satellites that survive longer than this minimum requirement -- constituting a subset of the original sample. The mean number of pericenters survived as a function of stellar mass (for the original sample) is displayed in the second to bottom panel of Fig. \ref{fig:s4:mstar-orbprops}. We also note that given a number satellites will share the same host halo, the results for host centrals will be biased towards more massive haloes in each sample -- i.e., those that contain the most satellites (and likely, a relatively massive central galaxy). These values are simply provided for reference.  

The top row of panels illustrates the orbital radius of satellites relative to $R_{200}$ of their host haloes. As noted in relation to Fig. \ref{fig:s4:mstar-orbprops}, we see that pericentric and apocentric radii tend to decrease with stellar mass. Additionally,  we see that apocentric radii decrease with the number of orbits completed for our selection -- which we attribute to the effects of dynamical friction \citep{DeLucia2012}. Across stellar mass, the  orbital radii between the two halo mass samples are statistically similar.

The second row of panels illustrates ISM mass (gas with $T\leq5\times10^{4}\ {\rm K}$ within $R_{\rm BMP}$) as a function of orbital time. ISM masses of zero are set to a floor value of $10^6 {\rm M}_{\odot}$, slightly less than the mass of a single gas particle. Focusing on intermediate-mass satellites \mstar$\in[9,10]$, we observe that the average satellite in this bin begins with a pre-infall ISM mass just below $\approx10^{10}{\rm M}_{\odot}$. Full depletion of the ISM is normally complete by first apocenter in the large group/cluster sample, while, on average, it takes until second apocenter to fully deplete the satellites in the group-mass sample. Considering the high-mass satellites in the right-hand column, \mstar$\in[10,11]$, we see more protracted depletion times; with satellites in the large group/cluster sample normally depleting all of their pre-infall gas (just above $\approx10^{10} {\rm M}_{\odot}$) by second pericenter, and those in the group-mass host sample able to retain a gas mass of $> 10^{9} {\rm M}_{\odot}$ even after second pericenter. Host central ISM mass, on average, shows a very slight decrease with orbital time for each sample, consistent with internally-driven gas consumption.

The third row of panels illustrates CGM mass (hot gas with $T>5\times10^{4}\ {\rm K}$ associated with the satellite subhalo) as a function of orbital time. Focusing first on the intermediate-mass satellites, we observe that the onset of CGM stripping occurs quite early in all samples, at around $(t-t_{\rm fp})/T_{\rm orb}\approx-0.5$ or $r_{\rm orb}/R_{200}\approx2$. Full CGM removal is expected for the large group/cluster sample at first pericenter, and for the smaller group sample at second pericenter. While all hot gas, on average, is expected to be stripped by second pericenter for intermediate-mass satellites the smaller group sample -- we see this is not necessarily the case for the ISM reservoir, which is removed after some delay.

Considering the high-mass satellites, we note that the average CGM mass in this sample drops by a full $1.5$ dex ($10^{10}{\rm M}_{\odot}$ to $10^{8.5}{\rm M}_{\odot}$) from $(t-t_{\rm fp})/T_{\rm orb}\approx-1$ to first pericenter, however the hot gas is typically only fully depleted after a subsequent pericenter in the larger group/cluster sample, or reduced to $\approx10^{8}{\rm M}_{\odot}$ in the small group sample. Between the stellar mass samples, it is clear that CGM hot gas stripping from satellites on infall {\it precedes} cool gas stripping. This agrees with the idea that the more tenuous, hot gaseous haloes surrounding infalling satellites are relatively easy to deplete with ram pressure given their increased spatial extent and reduced density (e.g \citealt{Larson1980,Balogh2000a,McCarthy2008,Bahe2013}). Finally, we also note the growth of host central CGM mass to, on average, $10^{12}{\rm M}_{\odot}$ over the orbital times shown ($\approx10$ Gyr) -- likely due to a combination of IGM accretion and satellite stripping \citep{Angles-Alcazar2017,Mitchell2020b}. 

The fourth row of panels in Fig. \ref{fig:s4:trel-orbprops} shows average specific star formation rates as a function of orbital time (i.e. the mass of new stars formed within $R_{\rm BMP}$ over the designated time interval), divided by stellar mass. For galaxies with a measured average star formation rate of zero, we set the sSFR to a floor value of $10^{-13}\ {\rm yr}^{-1}$. Prior to infall, each of the satellite samples show increased specific star formation rates ($>10^{-10}\ {\rm yr}^{-1}$) compared to the host central sample. This is a simple reflection of declining {\rm sSFR} with stellar mass, which exists in the absence of environmental effects. We find that the departure of galaxies from their expected secular {\rm sSFR} upon infall to a host is dependent on the mass of the satellite, and the mass of the host halo. In general, this departure very closely follows the depletion of ISM mass, as shown in the second row of columns (discussed above).

Interestingly, for satellites above $10^{9} {\rm M}_{\odot}$ in the small group sample, \mhalo$\in[12.5,13.5)$, we measure the specific star formation rates to be very similar to the central galaxies even after a second pericenter. While the rate of star formation decline is steeper in these satellites than for the host centrals at this orbital time, the rate of change in star formation rate for these satellites is still quite slow -- dropping  $\approx0.5$ dex in sSFR over a full orbital time-scale, compared to the host centrals where sSFR drops by $\approx0.25$ dex over the same period. This agrees with empirical inferences of quenching occurring several Gyr after infall in low-mass groups (e.g. \citealt{Wetzel2013,Wheeler2014}).

The fifth and sixth row of panels in Fig. \ref{fig:s4:trel-orbprops} illustrate the excess specific ISM inflow ($\Delta \dot{M}_{\rm in}/M_{\rm ISM}$) and outflow ($\Delta \dot{M}_{\rm out}/M_{\rm ISM}$) rates in satellites along their infall. Here, the term {\it excess} means that we have normalised the specific inflow and outflow rates by the flow rates experienced by central galaxies, controlled for their stellar mass and redshift -- i.e. the flow rates ``expected'' in the absence of environmental effects. These expected flow rates were illustrated in Fig. \ref{fig:s3:mstar-flowspec-mhalo_z0} as a function of stellar mass. 

Firstly, we note that inflow rates to the ISM of galaxies follow the behaviour of the CGM hot gas reservoir -- i.e. the origin of this accreting gas. In both satellite samples, we observe a delay in the onset of inflow suppression {\it subsequent} to the onset of hot gas stripping. While hot gas begins to be stripped at $(t-t_{\rm fp})/T_{\rm orb}\approx-0.5$ or $r_{\rm orb}/R_{200}\approx2-2.5$, this decrease in available gas to feed the galaxy only significantly influences inflow rates much closer to pericenter (well within $R_{\rm 200}$) in these samples. 

We find that intermediate-mass satellites in large groups/clusters, on average, are completely cut-off from the accretion of new gas after first pericenter, while it requires 2 pericentric passages to achieve the same effect in a satellite of the same mass in a smaller group environment ($<10^{13.5}{\rm M}_{\odot}$). These findings match the typical time it takes to strip these satellite's CGM gas reservoirs. In the high-mass satellite sample, we find that satellites on infall to large groups/clusters, on average, experience a reduction in accretion rates relative to ISM mass of 1 dex after first pericenter, and are completely cut-off from accretion after a second pericenter. Interestingly, it appears that higher-mass satellites in the smaller group sample, \mhalo$\in[12.5, 13.5)$, experience a relatively mild decrease in specific inflow rates (at maximum $\approx0.3-0.4$ dex relative to the value expected for centrals -- remaining steady at this value after first pericenter). 

Concentrating on the excess outflow rates presented in the sixth row of panels in Fig. \ref{fig:s4:trel-orbprops}, we can see clear signatures of cool gas stripping along the orbit of satellites. Prior to infall, specific outflow rates in each sample are relatively uniform, sitting at a value slightly less than that expected in centrals. In each of these samples, we note a significant increase in outflow rates relative to centrals of $\approx0.2-0.4$ dex approaching pericenter. We argue that this corresponds to the influence of ram pressure stripping, which is most prevalent at pericenters where orbital velocities peak ($P_{\rm ram}\propto v_{\rm orb}^2$). The peak in excess gas outflow is the greatest for intermediate mass satellites in large groups/clusters ($\log \Delta \dot{M}_{\rm in}/{M_{\rm ISM}}\approx0.4$, $2-2.5$ times that experienced by central galaxies of the same mass) and is smallest in high-mass satellites in smaller group environments ($\log \Delta \dot{M}_{\rm in}/{M_{\rm ISM}}\approx0.15$, enhanced $40\%$ relative to central galaxies of the same mass). We measure that the peak in excess outflow rates appears to occur just subsequent to first pericenter for both samples, however we remind the reader that the gas flux measured at a given time corresponds to the flow of gas that occurred over the 2 snipshots prior. Between pericenters, outflow rates reduce to slightly above that of centrals for intermediate-mass satellites, and to a level similar to centrals in high-mass satellites.  

Finally, the bottom row of panels in Fig. \ref{fig:s4:trel-orbprops} shows the gas-phase metallicity of {\it star-forming gas}\footnote{We only include star-forming gas to calculate metallicities, as this better corresponds to gas-phase metallicities measured in observations \citep{Bahe2017}.} in the ISM of satellites as a function of orbital time. The metallicity of galaxies is often used as a proxy for the age and/or star formation history of galaxy populations over time, with recent observational and theoretical literature indicating that satellites in higher-density environments exhibit a systematically increased (gas-phase) metallicity compared to the field (e.g. \citealt{Cooper2008a,Dave2011,Pasquali2012,DeRossi2015,Genel2016,Bahe2017}). 

As fully explored in \citet{Schaye2015}, we remark that the mass-metallicity relation in \eagle\ does not match observations below $M_{\star}\approx10^{10}{\rm M}_{\odot}$ at fiducial resolution. We note that here, however, our results focus on the {\it relative} change in ISM metallicity of satellites over their orbit -- which we expect to be less sensitive to resolution effects.  

Host central galaxies show a very gradual enhancement in metallicity over orbital time, increasing $\approx0.1$ dex over the full time range shown and hovering near solar metallicity ($Z_{\rm ISM}\approx Z_{\odot}$) -- consistent with internally driven enrichment from stellar evolution and feedback. For each sample of satellites, the increase in metallicity over the same time interval is significantly larger than their host centrals, ubiquitously $>0.4$ dex by second pericenter. Pre-infall metallicity differs between intermediate-mass satellites and high-mass satellites, beginning at $\log(Z_{\rm ISM}/{\rm Z}_{\odot})\approx0$ and $\log(Z_{\rm ISM}/{\rm Z}_{\odot})\approx0.2$ respectively. Both intermediate and high-mass satellites start to see significant increases in metallicity prior to first pericenter -- $(t-t_{\rm fp})/T_{\rm orb}\approx-0.5$ -- as the galaxy is starved of low-Z inflow.

In both intermediate and high-mass satellites, we find that ISM metallicities reach a maximum plateau value by first apocenter -- $(t-t_{\rm fp})/T_{\rm orb}\approx0.5$. In the small group sample, this plateau value sits at $\log(Z_{\rm ISM}/{\rm Z}_{\odot})\approx0.6$, compared to the large group/cluster sample (host to more severe ISM stripping) at $\log(Z_{\rm ISM}/{\rm Z}_{\odot})\approx0.8$. The absolute change in metallicity is larger for the intermediate mass satellites, which start with solar ISM metallicities compared to the high-mass sample, that show slightly super-solar pre-infall metallicities. This metallicity jump relative to field galaxies is higher than that typically quoted in observations, however, we note that the increase we observe is restricted to our sample of satellites that have definitively undergone an $R_{200}$ crossing and first pericentric passage.

Typically, gas that is radially further from the center of a galaxy exhibits lower metallicity due to the presence of freshly accreted low-$Z$ gas, and lower star formation-rate density (e.g. \citealt{Bahe2017,Collacchioni2020}). Thus, the preferential removal of the less bound, low-$Z$ gas in satellites leads to an increase in the metallicity of the ISM, as the measurement becomes constricted to the more central regions. This is compounded by the lack of low metallicity gas inflow, which otherwise plays a key role in modulating the metallicity of the ISM \citep{Collacchioni2020}. 

We note that while the relatively prolonged decrease in sSFR is clearly linked with decreasing ISM gas mass in satellites along their orbits, the rapid increase in gas-phase metallicity appears to be associated with the immediate removal of CGM gas at first-infall, and consequent lack of low-metallicity inflow. Our results also support the idea that the steep increase in gas-phase metallicity of satellites at first pericenter could be taken into account when inferring whether a satellite is still on first infall to a group, or has indeed experienced its first pericentric passage.

Holistically speaking, these findings support the conclusion that the more massive a satellite, the greater its propensity to retain a star-forming, cool gas reservoir (e.g. \citealt{Hester2006}). Broadly, this prediction aligns with the observation that dwarf satellites within the Milky Way's virial radius are quenched, while some satellites with \mstar$>10$ in the Virgo cluster retain a significant amount of gas  -- even despite the latter environment having significantly higher density (and thus ram pressure stripping potential) compared to the Local Group \citep{Cortese2021}. This idea also agrees with the theoretical findings of \citet{Bahe2015} and \citet{Oman2021}, who find that the transition time-scales of satellites to being passive after infall are significantly shorter ($< 500$ Myr) for satellites below $M_{\star}\approx10^{10}{\rm M}_{\odot}$ compared to their higher mass counterparts using the {\sc GIMIC} \citep{Crain2009} and Hydrangea simulations respectively \citep{Bahe2017b}. The exact time-scale for quenching depends heavily on when the ``zero'' time is taken, which is commonly set to ``infall'', or first pericenter \citep{Cortese2021}. 

\citet{Wright2019} also show that the quenching time-scales of \eagle\ satellites are significantly shortened ($<1$ Gyr) for galaxies with small stellar--halo mass ratios ($M_{\star}/{\rm M}_{\odot}<10^{-4}$; i.e. low-mass satellites in higher-mass haloes). Our findings suggest that some satellites can be fully quenched after just one pericentric passage -- which we indeed find to be the expectation for satellites with stellar mass less than $10^{10}{\rm M}_{\odot}$ on infall to haloes above a mass of $10^{13.5}{\rm M}_{\odot}$; and even for satellites in less massive group haloes, between $10^{12.5}{\rm M}_{\odot}$ and $10^{13.5}{\rm M}_{\odot}$ when their stellar mass is less than $10^{9}{\rm M}_{\odot}$.

Fig. \ref{fig:s4:trel-ormemclass_mstar} uses the gas flow classes outlined in Table \ref{tab:s3:emclass} with average inflow, outflow and star formation rates shown in Fig. \ref{fig:s4:trel-orbprops} to plot the changing evolutionary state of our different satellite samples along their orbits. We use the same satellite sample classifications (based on stellar and halo mass) as employed for Figs. \ref{fig:s4:mstar-orbprops} and \ref{fig:s4:trel-orbprops}.  We also remind the reader that some pre-infall galaxies may indeed be technically classified as a central, with the requirement for selection in the satellite sample only enforced on satellite/central status at first pericenter. 

Focusing on intermediate mass satellites, \mstar$\in[9,10)$, we observe a more gradual transition to quiescence over their orbits in both halo mass samples. Pre-infall quenched fractions sit very close to zero -- growing to $\approx5\%$ at pericenter in the small group sample, and $\approx50\%$ at pericenter in the large group/cluster sample. At second pericenter, the quenched fraction grows to $\approx40\%$ in the small group sample, and $\approx80\%$ in the large group/cluster sample. In the small-group sample (middle column, top panel), we note that the fraction of galaxies net growing their gas reservoir remains just above $5\%$ over the first orbit. In the same sample, the fraction of galaxies with rapid depletion time-scales is largest just after first pericenter ($\approx60\%$) until apocenter, where the fraction of galaxies with longer predicted depletion time-scales ($>1.5$ Gyr) grows to $\approx30\%$. Based on the results in Fig. \ref{fig:s4:trel-orbprops}, it appears that some galaxies in this sample (intermediate mass satellites in the small-group selection) are not completely cut-off from gas inflow after first pericenter, and at first apocenter, outflow rates reduce by a factor of a few relative to their value at pericenter. The combination of these factors means that such galaxies deplete their gas reservoirs relatively slowly -- while inflow rates are not enough to fully replenish the ISM, the lack of stripping at apocenter slows the quenching process. 

On approach to second pericenter, the fraction of quenched galaxies continues to increase and galaxies with slow depletion time-scales move to the fast-depletion group as outflow rates again increase. In the large group/cluster sample (middle column, bottom panel), we note that the likelihood of finding a slow-quenching galaxy at first apocenter is much lower ($<10\%$) than found for the small group sample. Referring to Fig. \ref{fig:s4:trel-orbprops}, we find that galaxies in this sample are fully cut off from fresh accretion after first pericenter, associated with more efficient stripping of the satellite's CGM gas in the harsher group/cluster environment. We do not necessarily find there to be more severe specific ISM outflow rates compared to the less massive group sample, however the lack of replenishing gas means the quenching process is greatly accelerated for these galaxies between first and second pericenter.

The picture for high-mass satellites involves a slightly slower growth rate of the quenched population along their orbits. As per the ${M}_{\star}$--sSFR relation for central galaxies, the ${\rm sSFR}$ of these high-mass satellites has a pre-infall value slightly below the lower mass groups, and a correspondingly larger quenched fraction of $\approx10\%$ compared to $<5\%$. At first pericenter, this value remains similar at $\approx10\%$ in the small group sample, and grows to $\approx15-20\%$ in the large group/cluster sample. At second pericenter, the quenched fraction grows to $\approx35\%$ in the small group sample, and $\approx80\%$ in the large group/cluster sample. While these second pericenter quenched fractions are similar to the intermediate-mass sample, the growth of the quenched fraction is slightly gentler given the higher pre-infall quenched fractions in this sample. We note a clear increase in the fraction of galaxies in the large group/cluster sample (right-most bottom panel) that are part of the slowly quenching group near first apocenter ($\approx15\%$), compared to this fraction in the equivalent intermediate-mass sample ($<10\%$). We remark that this is likely due to the propensity of massive satellites to retain a CGM reservoir after first pericenter, and as such, still experience some level of ISM accretion (refer to Fig. \ref{fig:s4:trel-orbprops}). This is not the case for the intermediate-mass sample, which, on average, experiences more dramatic hot gas stripping (and thus starvation) after first pericenter. 

The results presented in Figs. \ref{fig:s4:trel-orbprops} and \ref{fig:s4:trel-ormemclass_mstar} highlight the importance of considering the {\it interplay} between gas inflows (or a lack thereof), together with hot and cold gas removal in satellites; and moreover, how these processes {\it vary temporally} along the orbit of these galaxies in driving their path towards quiescence in group environments. This path towards quiescence will typically begin with the removal of hot gas along first-infall, which becomes evident at $2-3$ virial radii and becomes more efficient with increasing host mass, decreasing satellite mass, and increasing orbital velocity (which is periodic, peaking at pericenters, and reducing at apocenters). Once significant hot gas removal has occurred, the galaxy will experience ``starvation'' -- with no surrounding hot gas left to replenish the ISM reservoir. The depletion of this hot gas reservoir also subsequently allows direct cold gas (ISM) stripping, which also demonstrates periodic behaviour prior to complete depletion. In the absence of CGM gas, the efficiency of cold gas stripping is a strong function of host mass and satellite mass -- with low-mass satellites in large group/cluster environments experiencing very efficient stripping in the absence of accretion, and higher-mass satellites experiencing less efficient stripping and a more gradual, starvation-like quenching scenario after their first pericenter. 

Accounting for each of the above factors, our analysis leads to the following predictions for the quenching of satellites along their orbits in a group environment (with ``quenching'' defined by a threshold of $\log({\rm sSFR}/{\rm yr}^{-1})<-11+0.5z$). We predict that the average intermediate-mass satellite, \mstar$\in[9,10]$, will be quenched subsequent to second pericenter in a small group environment; and will be quenched just after first pericenter in a large group/cluster environment. Secondly, we predict that $\approx30\%$ of high mass satellites, \mstar$\in[10,11)$, will be quenched after second pericenter in a small group environment; and that the average high-mass satellite will be quenched near first apocenter in a large group/cluster environment. The rate of growth of the quenched fraction is gradual in many cases, and as such, these orbital quenching predictions represent the averages within some spread. 


\section{Discussion and Summary}\label{sec:conclusions}

In this paper, we analyse the balance of gas flows in \eagle\ satellite galaxies as a function of environment (\S\ref{sec:results:mhalo}) and orbital time (\S\ref{sec:results:orbits}). To do so, we employ a robust methodology to define the ISM of \eagle\ galaxies, and make use of the temporally finely-sampled \eagle\ ``snipshots'' to fairly quantify and compare ISM inflow, outflow, and star formation rates. 

In \S\ref{sec:results:mhalo}, we quantify how the gas inflow, outflow, and star formation rates compare in central and satellite galaxies as a function of environment in the context of the equilibrium model. For galaxies experiencing net ISM depletion, we can also use these flow rates to calculate an indicative ISM depletion time-scale. We find that:

\begin{itemize}
    \item There is clear, environmentally-driven suppression in the gas accretion rates of satellite galaxies, becoming increasingly significant with increasing host halo mass (agreeing with the findings of \citealt{Voort2017}). 
    
    \item When ISM gas mass is controlled for, the rate of specific gas outflow (accounting for both internal gas removal mechanisms and externally-driven stripping) increases with increasing halo mass in satellite galaxies. 
    
    \item Central galaxies typically stay very close to equilibrium at redshifts $z<2$ -- that is to say, inflow rates balance star formation and outflows. For those galaxies that do experience net gas depletion, their expected depletion time-scales are longer ($>1.5$ Gyr) than satellites. 
    
    \item Satellites experience increasing net gas depletion with increasing halo mass, with many exhibiting expected depletion time-scales below $1$ Gyr.
    
    \item While the lack of gas inflow to satellites in large group/cluster mass haloes could completely explain their quenching, any sub-Gyr quenching time-scales require concurrent cold gas removal. 
\end{itemize}

In \S\ref{sec:results:orbits}, we make use the \orbweaver\ code to quantify the orbits of satellites in the \eagle\ simulations. We use this orbital catalogue to analyse the properties of satellites as a function of orbital time, defined by the time of first pericentric passage and normalised by orbital time-scale. With our orbital sample, we conclude that:

\begin{itemize}

    \item Satellites begin their path to quiescence with the stripping of circum-galactic hot gas at $2-3$ virial radii from the host. This hot gas removal is most efficient in large group and cluster environments ($M_{200}>10^{13.5}M_{\odot}$), and takes longer for high-mass satellites ($M_{\star}>10^{10}M_{\odot}$). 
    
    \item The removal of this hot gas is clearly correlated with (i) the onset of starvation -- that is, a lack of gas inflow to the ISM; and (ii) the onset of cold gas stripping, where there is no longer a ``buffer'' between the ISM and surrounding intra-group/intra-cluster medium. 
    
    \item In low-mass satellites, \mstar$<10$, the suppression of gas inflow is typically permanent after hot gas stripping, which normally occurs at first pericenter. In contrast, some high-mass satellites, \mstar$>10$, can maintain a small hot gas reservoir and experience continued ISM gas accretion after first pericenter.
    
    \item The efficiency of cold gas stripping is periodic, peaking at pericenters where the relative velocity between the satellite and intra-group/intra-cluster medium (and thus ram-pressure force, $P_{\rm ram}\propto v^{2}$) is maximised. 
    
    \item The gas-phase metallicity of satellites in all mass samples increases dramatically (by $>0.5$ dex) around first pericenter. After this steep jump at first pericenter, gas metallicities increase relatively slowly. 
    
    \item The prolonged reduction in SFR along the orbit of a satellite is closely-linked with ISM gas removal via outflows, while the relatively rapid removal of CGM gas at first-infall -- and consequent lack of low-Z gas inflow to the ISM -- appears to be linked with the rapid increase in gas-phase metallicity.
    
\end{itemize}

We also produce predictions for the expected quenching of satellites as a function of their orbital time, split by stellar and halo mass:

\begin{itemize}
    \item The average intermediate-mass satellite, \mstar$\in[9,10)$, will be quenched subsequent to a second pericenter in a small group-mass environment -- \mhalo$\in[12.5,13.5)$; and will be quenched just {\it after} first pericenter in a large group/cluster environment -- \mhalo$\in[13.5,15]$.
    \item Only $\approx 30\%$ of high-mass satellites, \mstar$\in[10,11)$, will be fully quenched even after a second pericenter; while the average high-mass satellite will be quenched near first apocenter in a large group/cluster environment -- \mhalo$\in[13.5,15]$.
    
\end{itemize}
We remind the reader that our sample represents satellites that have largely experienced their first pericenter prior to $z=0.5$, and remark that the evolution of satellites on first-infall at $z\approx0$ is likely to be slightly different. Given that $z\approx0$ galaxies experience globally lower gas accretion rates and possess lower initial gas-fractions, it is likely that ISM depletion may occur over shorter time-scales in group environments

By quantifying the different gas flows and processes involved in the quenching of satellites along their orbits, our findings begin to break the degeneracy between different quenching mechanisms. Our results highlight the fact that both starvation -- a lack of gas inflow; and cold gas stripping -- direct gas removal from the ISM, occur simultaneously in satellites. The magnitude of the influence of each mechanism is strongly dependent on the way in which one defines each process -- however we argue that each plays a comparable and necessary role in the eventual quenching of satellites. 


\section*{Acknowledgements}
 RW is funded by a Postgraduate Research Scholarship from the University of Western Australia (UWA). CL is funded by the ARC Centre of Excellence for All Sky Astrophysics in 3 Dimensions (ASTRO 3D), through project number CE170100013. CL also thanks the MERAC Foundation for a Postdoctoral Research Award. CP acknowledges the support ASTRO 3D. ARHS is grateful for funding through the Jim Buckee Fellowship at ICRAR/UWA. LC acknowledges support from the Australian Research Council Discovery Project and Future Fellowship funding schemes (DP210100337 and FT180100066). This work made use of the supercomputer OzSTAR, which is managed through the Centre for Astrophysics and Supercomputing at Swinburne University of Technology. This super-computing facility is supported by Astronomy Australia Limited and the Australian Commonwealth Government through the National Collaborative Research Infrastructure Strategy (NCRIS). The \eagle\ simulations were performed using the DiRAC-2 facility at Durham, managed by the ICC, and the PRACE facility Curie based in France at TGCC, CEA, Bruyeres-le-Chatel. 

The authors used the following software tools for the data analysis and visualisation in the paper: 
\begin{itemize}
    \item {\fontfamily{pcr}\selectfont python3} \citep{VanRossum1995}
    \item {\fontfamily{pcr}\selectfont numpy} \citep{Harris2020}
    \item {\fontfamily{pcr}\selectfont scipy} \citep{Virtanen2020}
    \item {\fontfamily{pcr}\selectfont matplotlib} \citep{Hunter2007}
\end{itemize}

\section*{Data Availability}

Particle data from the set of the \eagle\ runs used for our analysis are publicly available at \url{http://dataweb.cosma.dur.ac.uk:8080/eagle-snapshots/}. Galaxy gas flow rate catalogs are available upon request to the corresponding author. 



\bibliographystyle{mnras}
\bibliography{references.bib}

\appendix
\newpage
\section{Defining the ISM and gas flows}\label{sec:app:gasflow}
Here we detail the methodology employed to define the ISM of \eagle\ galaxies, together with the relevant gas inflow, outflow, and star formation rates presented in this paper. As mentioned in \S\ref{sec:methods:gasflow}, for the purposes of this paper it was important for us to define the ISM of \eagle\ galaxies  in a manner as agnostic as possible to the galaxy's central or satellite classification. We focus on gas flows to and from the ``cool'' gas reservoir of galaxies, reflecting what could be considered the gas linked to a neutral molecular \citep{Lagos2015} or atomic \citep{Bahe2016} phase.

\subsection{Previous literature}\label{sec:app:gasflow:prev}
Here we provide an overview of the different ISM definitions that have been utilised in previous studies to analyse gas flow rates in cosmological hydrodynamical simulations. 

Focusing on gas inflow, \citet{Voort2017} define the ISM of both central and satellite galaxies in \eagle\ by selecting star-forming particles within a $30$kpc aperture of a galaxy's centre of potential. Gas particles with a non-zero SFR in \eagle\ reflect a ``cold'' phase, meeting the metallicity-dependent density threshold and being within $0.5$~dex of the density-dependent temperature floor outlined in \citet{Schaye2015}. In a similar manner, \citet{Correa2018b} define the ISM of \eagle\ central galaxies by selecting particles within a radial distance of $0.15\times R_{\rm 200}$ of the host centre of potential, that are either (i) star forming, or (ii) part of the ``atomic'' ISM (with $n_{\rm H}>10^{-1}{\rm cm}^{-3}$ and $T<10^{5}{\rm K}$). \citet{Mitchell2020b} take a slightly different approach in defining the ISM of \eagle\ central galaxies, including gas particles within $0.20\times R_{\rm 200}$ that are either (i) star forming, or (ii) meet an atomic-phase selection; selecting particles with $n_{\rm H}>10^{-2}{\rm cm}^{-3}$ that are within $0.5$~dex of the density-dependent temperature floor.

Each of these works quote accretion rates to galaxies as the mass flux of gas particles that were not part of the ISM of a galaxy at a snapshot $z_{\rm i}$, but do meet the ISM conditions at the subsequent snapshot $z_{\rm j}$. This mass flux is normalised by the time interval between snapshots $z_{\rm j}$ and $z_{\rm i}$, $\Delta t_{\rm ij}$. Instead focusing on outflows, \citet{Mitchell2020a} use the same ISM selection and simulation cadence as \citet{Mitchell2020b}, imposing an additional average velocity requirement to ensure genuine outward radial movement of outflow particles. In this case, the outflow rate is computed as the mass flux of particles which were part of the ISM of a galaxy at a snapshot $z_{\rm i}$, do not meet the ISM conditions at the subsequent snapshot $z_{\rm j}$, and also meet the aforementioned average radial velocity requirement. 

\citet{Voort2017} and \citet{Correa2018b} both implement a $\Delta t_{\rm ij}$ defined by the cadence of the $28$ \eagle\ snapshots (including $9$ outputs between $z=0$ to $z=1$, and $8$ outputs between $z=1$ and $z=2$); while \citet{Mitchell2020a} and \citet{Mitchell2020b} instead use a $\Delta t_{\rm ij}$ defined by $200$ \eagle\ ``snipshot'' outputs (including $65$ outputs between $z=0$ and $z=1$, and $36$ outputs between $z=1$ and $z=2$). The latter choice to use a finer time-cadence ensures that instantaneous outflow rates are converged, which can be artificially reduced by recycling material if a longer $\Delta t_{\rm ij}$ is used. 

\subsection{The ``BaryMP'' method}\label{sec:app:gasflow:ism}

\begin{figure}
\centering
\includegraphics[width=1\columnwidth]{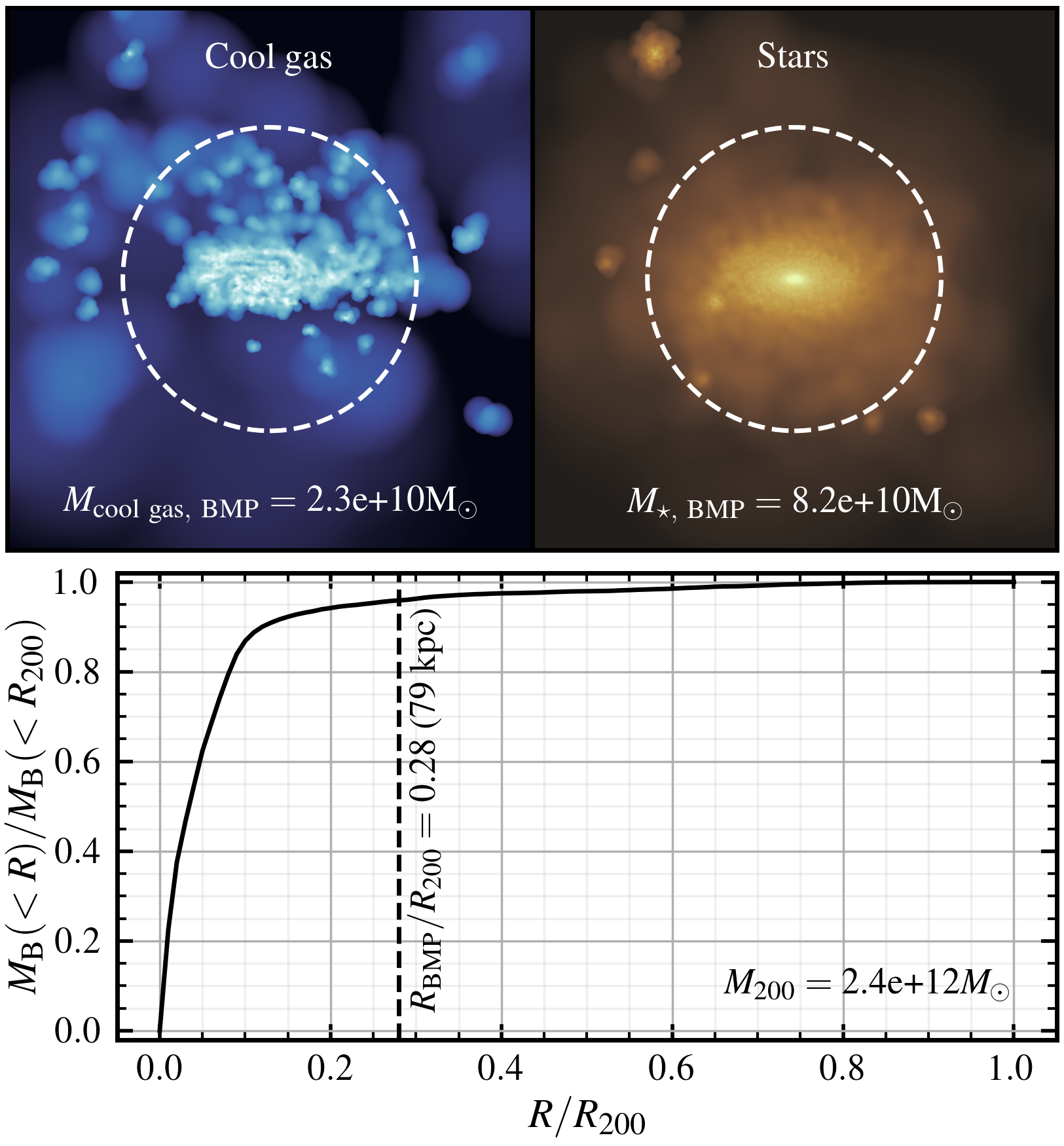}
\caption{The ``BaryMP'' (BMP) fitting process for a $z=0$ \mstar$\approx11$ galaxy (based on \citealt{Stevens2014}). The cumulative mass profile of cool gas (${\rm SFR}>0$ or $T<5\times10^{4}$ K; top left panel) and stars (top right panel) is fit to find the point at which the profile becomes effectively isothermal. In this case the BMP radius is found to be $79$ pkpc, as indicated by the dashed lines in each panel.}
\label{fig:s2:bmpfit}
\end{figure}

As outlined in \S\ref{sec:methods:gasflow} --  to define the extent of a galaxy in a manner agnostic to central/satellite status, we employ the baryonic mass profile (BMP) fitting method described in \citet{Stevens2014}. With this method, the cumulative radial baryonic mass profile of ``cool'' gas (particles with either $T<5\times 10^4$ K or non-zero SFR) and stellar particles is analysed for a given (sub)halo\footnote{We exclusively fit the baryonic mass profile using the particles associated with each \subfind\ (sub)halo. In the case of central galaxies, this means excluding nested subhaloes; and in the case of satellites, this means excluding the diffuse particles assigned to the central subhalo.} to identify the best-fit transition point in the profile where $\rho_{\rm bar}(r)\propto r^{-2}$. This is a good approximation for an isothermal diffuse hot gas halo, as well as for a DM halo in virial equilibrium at the scale radius \citep{Navarro1996}. Following the BMP fitting process, the ISM of a galaxy is then defined by all cool gas particles ($T<5\times 10^4$ K or non-zero SFR) internal to this radius ($R_{\rm BMP}$). Fig. \ref{fig:s2:bmpfit} demonstrates the BMP fitting process for a \mstar$\approx11$ \eagle\ galaxy at $z=0$.

\begin{figure}
\includegraphics[width=\columnwidth]{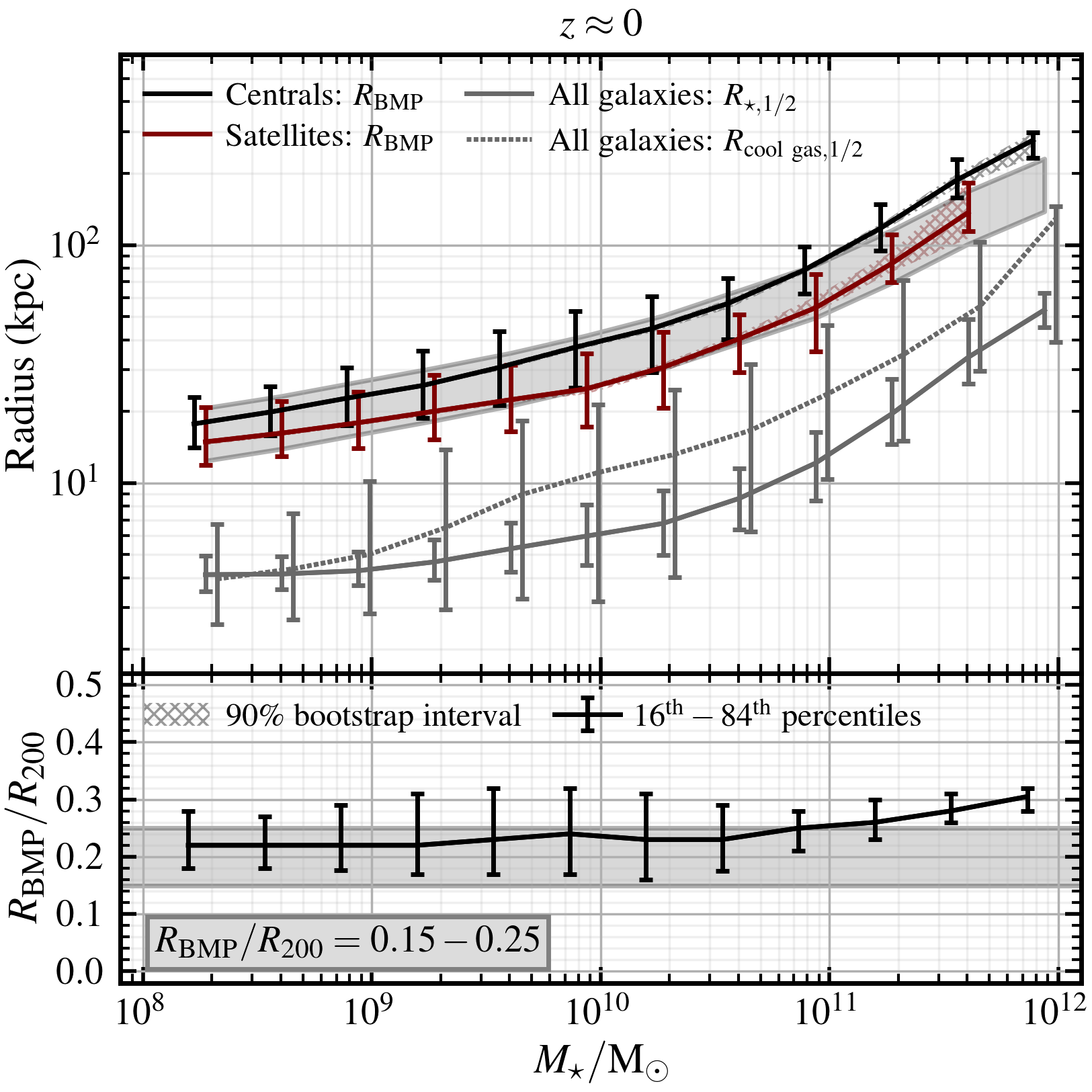}
\caption{Top panel: the median BMP radius, $R_{\rm BMP}$ as a function of stellar mass internal to this radius, $M_{\star,\rm\ BMP}$, for $z\approx0$ \eagle\ central (black) and satellite (red) galaxies. For reference, we also illustrate the stellar half-mass radius ($R_{\star,1/2}$, solid line) and cool gas half-mass radius ($R_{\rm cool\ gas,1/2}$, dotted line) for all galaxies in grey. Bottom panel: the median ratio of $R_{\rm BMP}$ to the host $R_{\rm 200}$ virial radius for $z\approx0$ central galaxies only. At fixed stellar mass, centrals exhibit systematically larger $R_{\rm BMP}$ values compared to satellites, by $\approx0.1$ dex on average.}

\label{fig:s2:bmpstats}
\end{figure}

Fig. \ref{fig:s2:bmpstats} demonstrates the results of the BMP fitting process for $z\approx0$ \eagle\ galaxies, split by central/satellite status. The top panel illustrates the median BMP radius, $R_{\rm BMP}$, as a function of stellar mass internal to this radius, $M_{\star,\rm\ BMP}$. Focusing on central galaxies (black), $R_{\rm BMP}$ increases with mass from a median of $25$ kpc at \mstar$\approx9$ to $300$ kpc at \mstar$\approx12$. The spread, as indicated by the grey shaded region, is largest for less massive systems, \mstar$<10.5$. In this regime, the $16^{\rm th}-84^{\rm th}$ percentile range remains at $\approx 0.2$ dex, while for more massive systems, \mstar$>11$, this spread reduces to $\approx 0.1$ dex.

In the top panel of Fig. \ref{fig:s2:bmpstats} we also illustrate the stellar half-mass radius ($R_{\star,1/2}$) and cool gas half-mass radius  ($R_{\rm cool\ gas,1/2}$) for all galaxies in grey. We note that these half-mass radii are calculated based on the particles within an aperture of $R_{\rm BMP}$. The reader may notice that while the half-mass radii follow the same functional form as the BMP radii with stellar mass, the $R_{\rm BMP}$ values (corresponding closely to $0.15-0.25\times R_{200}$) are significantly larger than $R_{\star,1/2}$. This highlights the fact that the BMP radius acts as an upper radial limit for the inclusion of particles in the stellar and ISM selection. In the case of cool gas, the temperature cut ($T<5\times10^{4}$ K) preferentially selects gas that is well within the $R_{\rm BMP}$ radius. Above $M_{\star}\approx10^{10}{\rm M}_{\odot}$, the half-mass radius of cool gas is a factor of $\approx2$ larger than the stellar half-mass radius, meaning that on average, our ISM selection corresponds to a gas distribution that is slightly extended relative to the stellar component. 

The bottom panel of Fig. \ref{fig:s2:bmpstats} shows the ratio of $R_{\rm BMP}$ to the host halo virial radius, $R_{\rm 200}$, for central galaxies. The median  $R_{\rm BMP}/R_{\rm 200}$ fraction ranges between $0.2-0.3$ for the majority of the mass range, exceeding 0.3 for the most massive galaxies, \mstar$>11.5$. These $R_{\rm BMP}/R_{\rm 200}$ values slightly exceed the radial cuts used for central galaxies in \citet{Correa2018b} and \citet{Mitchell2020b}. 

Comparing centrals and satellite galaxies in the top panel of Fig. \ref{fig:s2:bmpstats}, we note that median $R_{\rm BMP}$ values for central galaxies systematically exceed satellite $R_{\rm BMP}$ measurements by $\approx0.1-0.2$ dex ($25-50\%$) over the full mass range considered. Satellite galaxies with \mstar$\approx10-11$ exhibit higher variance in baryonic radii compared to centrals, with a $16^{\rm th}-84^{\rm th}$ percentile range of $\approx0.3$ dex. The more concentrated baryonic mass profiles associated with satellites are likely a reflection of their reduced gas content compared to centrals; in which case $R_{\rm BMP}$ is primarily set by the stellar component.

\bsp	
\label{lastpage}
\end{document}